\documentclass[a4paper,11pt]{article}
\pdfoutput=1 

\usepackage{jheppub} 

\usepackage[utf8]{inputenc}
\usepackage{slashed,cancel}
\usepackage{bm}
\usepackage{pifont}
\usepackage[english]{babel}
\usepackage[normalem]{ulem} 
\usepackage{booktabs}
\usepackage{braket}
\usepackage{MnSymbol}
\usepackage{graphicx}
\usepackage{subcaption}
\usepackage{multirow}
\usepackage[skins,theorems]{tcolorbox}
\usepackage[capitalise]{cleveref}
\usepackage{orcidlink} 
\tcbset{highlight math style={enhanced,
  colframe=red,colback=white,arc=0pt,boxrule=1pt}}
\usepackage{indentfirst}

\numberwithin{equation}{section}

\definecolor{rosso}{cmyk}{0,1,1,0.4}
\definecolor{rossos}{cmyk}{0,1,1,0.55}
\definecolor{rossoc}{cmyk}{0,1,1,0.2}
\definecolor{blu}{cmyk}{1,1,0,0.3}
\definecolor{blus}{cmyk}{1,1,0,0.6}
\definecolor{bluc}{cmyk}{1,1,0,0.1}
\definecolor{verde}{cmyk}{0.92,0,0.59,0.25}
\definecolor{verdec}{cmyk}{0.92,0,0.59,0.15}
\definecolor{verdes}{cmyk}{0.92,0,0.59,0.4}
\newcommand{\be}{\begin{equation}}
\newcommand{\ee}{\end{equation}}
\newcommand{\ba}{\begin{equation}\begin{aligned}}
\newcommand{\ea}{\end{aligned}\end{equation}}

\graphicspath{{arXiv_V2_Plots/}}

\title{\boldmath 
Multiple Axions in Laboratory Experiments
}

\preprint{IPPP/25/92, KEK-QUP-2025-0025, KEK-TH-2780}

\author[]{Arturo de Giorgi$^a$,}
\author[]{Joerg Jaeckel$^b$,}
\author[]{Sebastian Monath$^{b,c,d}$,}
\author[]{Volodymyr~Takhistov$^{e,f,g,h}$}

\affiliation[a]{Institute for Particle Physics Phenomenology, Department of Physics, Durham University,\\
 Durham DH1 3LE, U.K.}

\affiliation[b]{Institut f\"ur Theoretische Physik, Universit\"at
Heidelberg,\\
 Philosophenweg 16, 69120 Heidelberg, Germany}

\affiliation[c]{Université Lyon 1, CNRS, IP2I, UMR 5822, Villeurbanne, France}

\affiliation[d]{
École Normale Supérieure (ENS) de Lyon \\15 parvis René Descartes - BP 7000, 69342 Lyon Cedex 07, France}

\affiliation[e]{International Center for Quantum-field Measurement Systems for Studies of the Universe and Particles (QUP,WPI), High Energy Accelerator Research Organization (KEK),\\
 1-1 Oho, Tsukuba, Ibaraki 305-0801, Japan}

\affiliation[f]{Theory Center, Institute of Particle and Nuclear Studies (IPNS),\\
High Energy Accelerator Research Organization (KEK),\\
 1-1 Oho, Tsukuba, Ibaraki 305-0801, Japan}

\affiliation[g]{Graduate University for Advanced Studies (SOKENDAI),\\
1-1 Oho, Tsukuba, Ibaraki 305-0801, Japan}

\affiliation[h]{Kavli Institute for the Physics and Mathematics of the Universe (WPI),\\
Kashiwa Chiba 277-8583, Japan}

\emailAdd{arturo.de-giorgi@durham.ac.uk}
\emailAdd{jjaeckel@thphys.uni-heidelberg.de}
\emailAdd{sebastian.monath@stud.uni-heidelberg.de}
\emailAdd{vtakhist@post.kek.jp}

\abstract{
Axions and axion-like particles generically appear in extensions of the Standard Model.  While many searches assume only a single axion species, there may exist a whole spectrum of multiple such fields.
We develop general formulas for axion-photon oscillations in the presence of multiple axions and analyze the implications for experimental searches, including light-shining-through-a-wall experiments, helioscopes and haloscopes.  
We demonstrate that axion multiplicity can qualitatively alter observational signatures, particularly through coherence and interference effects. Multiple axions can not only enhance signals compared to single axion scenarios, but also suppress them. We show that variations of experimental parameters and searches allow identifying contributions of multiple axions and obtaining information about their properties.
}

\begin{document}
\maketitle
\flushbottom

\section{Introduction}

The discovery of the Higgs boson~\cite{CMS:2012qbp,ATLAS:2012yve} has underscored the important role of spin-0 fields in Nature. While the Higgs is scalar, pseudoscalars may play their part.
A well-known example of a pseudoscalar, the axion, has gone far beyond its origin as a solution to the strong CP problem of quantum chromodynamics (QCD)~\cite{Peccei:1977hh,Weinberg:1977ma,Wilczek:1977pj,Zhitnitsky:1980tq,Dine:1981rt,Kim:1979if,Shifman:1979if}.
Indeed, axions and axion-like particles (ALPs, whenever they are unconnected to the strong CP problem) are a compelling dark matter (DM)  candidate~\cite{Preskill:1982cy,Abbott:1982af,Dine:1982ah,Arias:2012az}. They are often realized as pseudo-Goldstone bosons of an underlying shift symmetry, making them naturally light particles, present in many extensions of the Standard Model (see below for examples pertinent to this paper). Consequently, significant experimental efforts are underway to explore the axion landscape (cf., e.g.~\cite{Kim:1986ax,Jaeckel:2010ni,Irastorza:2018dyq,Agrawal:2021dbo,Adams:2022pbo,Antel:2023hkf} for some reviews).
Note that, for brevity, from now on we will usually speak of axions, but this is meant to include both axions and ALPs.

It is also plausible to ask if one could have more than one pseudoscalar or axion. This is indeed motivated by theoretical developments
that draw attention to
scenarios in which multiple axions
appear. Prominent examples include the string theory-inspired ``axiverse''~\cite{Svrcek:2006yi,Arvanitaki:2009fg,Acharya:2010zx,Cicoli:2012sz,Bachlechner:2017zpb,Demirtas:2018akl,Halverson:2019cmy,Wang:2020gmi,Mehta:2020kwu,Cyncynates:2021xzw,Broeckel:2021dpz,Gendler:2023kjt,Dimastrogiovanni:2023juq,Agrawal:2024ejr,Leedom:2025mlr,Reig:2025dqb,Agrawal:2025rbr}, featuring a spectrum of axions that spans decades in masses and couplings, as well as extra-dimensional models with bulk axions where the four dimensional effective theory contains a tower of axion states with correlated masses and couplings following from the theory's geometric structure~\cite{Dienes:1999gw,Dienes:2011ja,Dienes:2011sa,deGiorgi:2024elx}. In both of these cases, multiple axions are a generic outcome of the underlying theory. Notably, the presence of multiple axions is compatible with a solution to the strong CP problem and can extend the allowed axion parameter space beyond that of the conventional QCD axion~\cite{Gavela:2023tzu}.

These considerations motivate broader experimental and observational probes. Scenarios with multiple axions have been extensively studied in the context of cosmology and astrophysics.
Notable studies include constraints on extra-dimensional axion dark matter~(DM)~\cite{Dienes:2011sa}, the impact on the cosmological evolution of topological defects~\cite{Benabou:2023npn,Lee:2024toz}, as well as general cosmological and DM probes~\cite{Mack:2009hs,Arvanitaki:2009fg,Acharya:2010zx,Luu:2018afg,Reig:2021ipa,Murai:2023xjn,Gendler:2023kjt,Glennon:2023jsp,Murai:2024nsp,Li:2025uwq,Dunsky:2025sgz,Dessert:2025yvk,Asadi:2025cvm,Haque:2025eyl}. Moreover, a variety of  astrophysical tests have been considered~\cite{DiLella:2000dn,Horvat:2001jy,Cicoli:2012sz,Chadha-Day:2021uyt,Bastero-Gil:2021oky,Chadha-Day:2023wub}.
However, the implications of multiple axions for laboratory searches and the interpretation of associated experimental signatures remain underexplored.\footnote{See, however,~refs.~\cite{Chadha-Day:2021uyt,Chadha-Day:2023wub} for a study also considering helioscopes~\cite{Sikivie:1983ip}.}

Therefore, in this work, we analyze manifestations of multiple axions in laboratory experiments, including light-shining-through-a-wall (LSW)~\cite{Okun:1982xi,Anselm:1985obz,Bibber} experiments, helioscopes and haloscopes~\cite{Sikivie:1983ip}.
We develop general formulas for axion-photon conversion in the presence of an arbitrary number of axions with a variety of masses and couplings, and derive the corresponding conversion probabilities relevant for laboratory searches. We demonstrate that multiple axions can qualitatively modify experimental signals. For $N$ sufficiently light and coherent axion fields, the resulting conversion probability can be enhanced by a factor of $\sim N^2$, while in incoherent regimes they typically scale linearly with $\sim N$.
As we show, additional axions do not necessarily lead to enhanced sensitivity as interference effects can instead suppress observable signals in fixed experimental configurations. We also discuss experimental strategies to probe multiple axions. These techniques enable laboratory searches to distinguish single and multi-axion scenarios and to obtain information about their underlying properties.

The paper is organized as follows. In section~\ref{sec:LSW-pheno}  we develop a general formalism, valid for the multi axion case, describing axion-photon propagation and conversion in the presence of magnetic fields. We discuss the role of coherence and interference effects, with emphasis on LSW experiments. In section~\ref{sec:models}, we discuss several representative benchmark models of multiple axions. Section~\ref{sec:LSW} examines the impact of multiple axions in LSW experiments and identifies experimental strategies to distinguish multi-axion scenarios from the single axion case. In section~\ref{sec:helioscopes} and~section~\ref{sec:haloscopes}  we discuss complementary searches for multiple axions in helioscope and haloscope experiments, respectively. We conclude in section~\ref{sec:conclusions}.

\section{Propagation and conversion in multi-axion systems}
\label{sec:LSW-pheno}

In the presence of external magnetic fields, axions can convert into photons and vice versa through their coupling to the electromagnetic field strength, via the so-called Primakoff effect. This conversion plays a central role in a broad class of axion laboratory searches. The existence of multiple axions can significantly affect such conversions. We first establish a general framework for axion-photon propagation and conversion in the presence of multiple axions. The resulting formalism forms the basis for the analysis of specific experimental configurations.

\begin{figure}
    \centering
    \includegraphics[width=\linewidth]{ 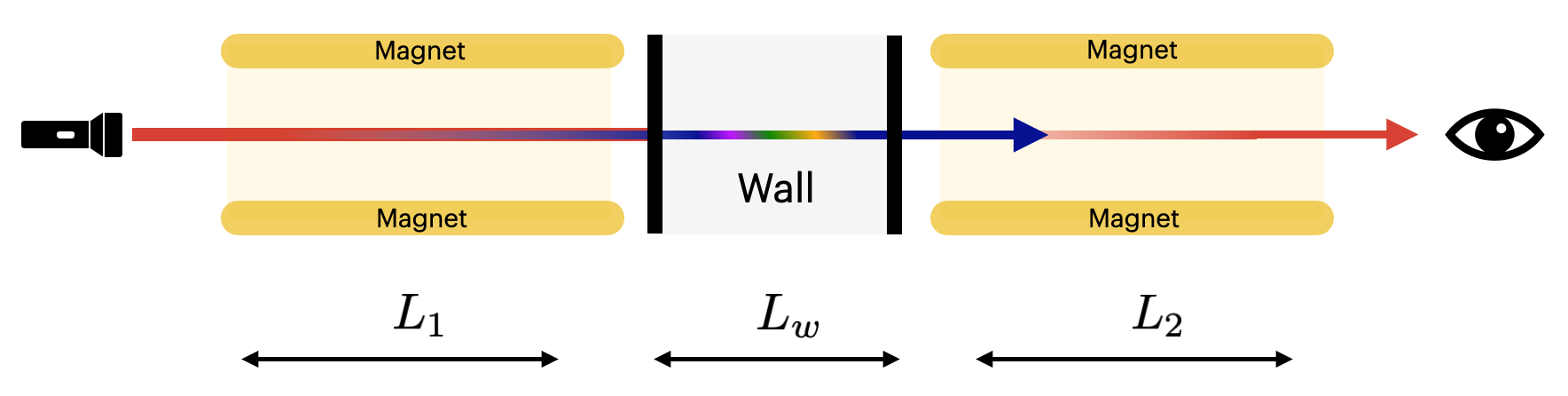}
    \caption{Schematic illustration of an LSW experiment. Photons from a laser beam (red)
    propagate through a magnetic field region of length $L_1$, where they can convert into a superposition of axion states through the axion-photon interaction.
    The axion state traverses an opaque to photons wall of thickness $L_w$. Subsequently, in a second magnetic field region of length $L_2$, axions can reconvert into photons that are~detectable.\label{fig:LSW-schematic}}
\end{figure}

We focus in particular on axion-photon conversion in LSW experiments, which allow for full experimental control over both axion production and detection. In an LSW setup, an incident laser beam traverses a magnetic field and produces axions. A wall blocks the photon flux while allowing the axion component to propagate through, after which axions can reconvert into photons in a second magnetic region and hence be detected. A schematic illustration is shown in figure~\ref{fig:LSW-schematic}. Within the magnetic field, the photon state is not a mass eigenstate and therefore undergoes oscillatory evolution before regeneration. In the presence of multiple axions, the same mechanism applies, but with a richer conversion structure arising from interference and coherence effects among the distinct axion states.

Let us consider  an  effective field theory Lagrangian describing $N$ axions coupled to  the  electromagnetic field
\begin{align}\label{eq:Lagr}
    \mathcal{L}_{\rm eff}&=\sum\limits_{n=1}^N \left(\frac{1}{2}\partial^\mu a_n\partial_\mu a_n-\frac{1}{2}m_n^2 a_n^2\right)-\frac{1}{4}F^{\mu\nu}F_{\mu\nu}-\frac{g_{a\gamma}}{4}\left(\sum\limits_{n=1}^N c_n a_n\right)F^{\mu\nu}\Tilde{F}_{\mu\nu}\,,
\end{align}
where  the  $a_n$ denote axion fields with masses $m_n$, $F_{\mu\nu}$ is the electromagnetic field strength tensor and $\tilde{F}_{\mu\nu}$ is its dual. Here, the coefficients $c_n$ parametrize the relative coupling strengths of the individual axions to photons, while the overall interaction scale is set by the axion-photon coupling $g_{a\gamma}$. Without loss of generality, we normalize by setting $c_1=1$.

In ultraviolet complete models, the coefficients $c_n$ and the masses $m_n$ can be expected to be correlated. A notable example are theories with extra dimensions where bulk axions lead to characteristic mass and coupling patterns (see e.g.\ refs.~\cite{Dienes:1999gw, deGiorgi:2024elx}).
In general, multiple axion systems can exhibit both kinetic and mass mixings. By suitable field redefinitions, the Lagrangian can be expressed in the above form.
Since our primary focus is on experimentally relevant effects rather than specific model details, we carry out our analysis directly in this canonical basis. Throughout this work, we further neglect potential axion self-interactions as well as interactions with fermions, as these do not affect axion to photon conversion processes in the laboratory experiments considered here.\footnote{Note, however, that additional couplings such as, e.g., an axion-electron coupling can add potential additional production and detection channels in some experiments. Further, additional interactions such as axion-fermion couplings can induce axion-photon couplings at loop level. Here, we consider that such effects are absorbed into the effective axion-photon coupling and do not affect conversion~probabilities.}

To illustrate the scaling behaviour of axion-photon conversion in the presence of multiple axions, we can consider the   interaction
\begin{align}
\mathcal{L}_{a\gamma}
= -\dfrac{g_{a\gamma}}{4}a  F^{\mu\nu}\tilde F_{\mu\nu}\,,
\qquad
a = \sum_{n=1}^N c_n a_n\,,
\end{align}
where the interacting field $a$ is a linear combination of all the mass eigenstates $a_n$. For simplicity, we assume that the coefficients  $c_n\simeq\mathcal{O}(1)$ are all of the same order of magnitude.
Then, the interaction state that is a coherent superposition of $N$ mass eigenstates is canonically normalised approximately by $1/\sqrt{N}$.\footnote{To understand the normalization we can consider a superposition of orthonormal states \begin{align*}
    &\ket{a}= \mathcal{N}\sum\limits_{n=1}^N c_n\ket{a_n}\,, &\braket{a_n|a_m}=\delta_{nm}\,,
\end{align*}
the normalization of $\ket{a}$ requires $\mathcal{N}^{-2}=\sum_{n=1}^N|c_n|^2=N$, where the last equality holds if $c_n=1$ for all~$n$.}
We can then write the interaction as
\begin{equation}
\label{eq:cohere1}
    \mathcal{L}_{a\gamma}=-\frac{1}{4}\left(\sqrt{N}g_{a\gamma}\right)\left(\frac{a}{\sqrt{N}}\right)F^{\mu\nu}\Tilde{F}_{\mu\nu} .
\end{equation}
Thus, the production amplitude for the normalised interacting state  $a$ scales as\footnote{Here, we denote by $a$ the specific linear combination of axion fields that interacts with magnetic~fields.}
\begin{align}
\label{eq:cohere2}
    \mathcal{M}_\text{prod}\propto \sqrt{N}g_{a\gamma} ,
\end{align}
with an analogous scaling for the conversion amplitude at detection, such as in the case of LSW experiments.

Between production and detection, however, the interacting state $a$ can oscillate into orthogonal axion combinations that do not couple to photons. The survival probability of the interacting state after propagating over a distance $l$ is governed by a length-dependent probability factor $P_{a\to a}(\ell)$ that accounts for coherence and interference effects among the mass eigenstates. Combining production and detection, the probability for photon regeneration after the wall in an external magnetic field, $\gamma+B\to \gamma+B$, scales as
\begin{equation}
\label{eq:estimation-sigma}
    P_{\gamma\to\gamma}(\ell)\propto (\sqrt{N}g_{a\gamma})^4\left(\frac{P_{a\to a}(\ell)}{P_{a\to a}(0)}\right)\propto  g_{a\gamma}^4  N^2\left(\frac{P_{a\to a}(\ell)}{P_{a\to a}(0)}\right) .
\end{equation}
For propagation lengths short compared to the coherence length, one has $P_{a\to a}(\ell)/P_{a\to a}(0)\simeq 1$, and the process is enhanced by a factor $N^2$ compared to the single-axion case. At larger distances where decoherence among the mass eigenstates suppresses the survival probability, one has $P_{a\to a}(\ell)/P_{a\to a}(0)\propto 1/N$ and the process scales as $N$. Thus,
\begin{equation}
\left(\frac{P_{a\to a}(\ell)}{P_{a\to a}(0)}\right) \propto
\begin{cases}
1 \quad\,\,\Rightarrow P_{\gamma\to\gamma} \propto N^2, &\quad \text{coherent regime},\\
1/N \Rightarrow P_{\gamma\to\gamma} \propto N, &\quad \text{decoherent regime}.
\end{cases}
\end{equation}
Such a behaviour was identified in the context of axions in extra-dimensional models in ref.~\cite{Dienes:1999gw} and investigated in detail for five-dimensional bulk QCD axions in ref.~\cite{Dienes:2011sa}. Closely related discussions have also appeared in neutrino physics in models with multiple sterile neutrino states~\cite{Dienes:1998sb,Long:2013ota,Fong:2017gke,deGiorgi:2025xgp}.

We proceed to a general derivation of the axion-photon conversion probability in the presence of multiple axions.
When relevant to the experiments we have in mind, we also include medium effects for photons, including refraction and absorption.

\subsection{Eigenstates in magnetic field backgrounds}

Let us begin considering the conversion of photons into a single axion species in the presence of a background magnetic field. For this, we take the  propagating  photon perturbation  $ F^{\mu\nu}_\gamma$ on top of a background magnetic field $F^{\mu\nu}_{B}$~\cite{Raffelt:1987im},
\begin{equation}    F^{\mu\nu}=F^{\mu\nu}_{B}+ F^{\mu\nu}_\gamma .
\end{equation}
The equations of motion (EoM) imply that only the photon polarization parallel to the background magnetic field,  $A_{||}$,  interacts with the axion. To leading order in the axion-photon coupling, this reads~\cite{Raffelt:1987im}
\begin{align} \label{eq:alpmixing}
    &\Box \begin{pmatrix}
        A_{||}\\
        a
    \end{pmatrix}+\begin{pmatrix}
        0 &-\kappa\\
        -\kappa & m_a^2
    \end{pmatrix}\begin{pmatrix}
        A_{||}\\
        a
    \end{pmatrix}=0 , \qquad  \kappa\equiv g_{a\gamma}\omega B\equiv \omega \beta,
\end{align}
where $\Box$ denotes the d'Alembert operator, $\omega$ denotes the energy and $B = |\vec{B}_{\perp}|$ is the magnitude of the magnetic field perpendicular to the propagation direction of the photon. When propagation occurs in a medium rather than in vacuum, photon propagation is modified by an effective mass,
\begin{equation}
m_\gamma^2\equiv \omega^2(1-n^2)\simeq 2\omega^2(1-n)\,,
\end{equation}
where $n$ is the refractive index of the medium, and the approximation holds for most gases relevant for experiments of interest with $n\simeq 1$.
A more general treatment, including absorptive effects, will be provided below.

We employ a simplified plane wave ansatz propagating along the $x$-direction in time $t$ for the fields $A_{||}(t,x),a(t,x)\propto e^{-i(\omega t-k x)}$, with momentum $k$. A more systematic derivation based on the eikonal approximation, as employed by ref.~\cite{Adler:2008gk}, can be found in appendix~\ref{app:eikonal}. The ansatz leads to the EoMs in medium
\begin{align}
    &\left[(-\omega^2+k^2)+\begin{pmatrix}
        m_\gamma^2 &-\kappa\\
        -\kappa & m_a^2
    \end{pmatrix}\right]\begin{pmatrix}
        A_{||}\\
        a
    \end{pmatrix}=0 .
\end{align}
Defining $\lambda^2\equiv \omega^2-k_\lambda^2$,  where $k_{\lambda}$ denotes the wave number of a propagation mode $\lambda$, non-trivial solution exists when
\begin{equation}
    \det\begin{pmatrix}
    m_\gamma^2-\lambda^2 &-\kappa\\
    -\kappa & m_a^2-\lambda^2
    \end{pmatrix}=0 .
\end{equation}
For a single axion species, this yields eigenvalues
\begin{align}\label{eq:single-axion}
    \lambda^2_\pm=\frac{1}{2}\left(m_a^2+m_\gamma^2\pm \sqrt{(m_a^2-m_\gamma^2)^2+4\kappa^2}\right)\simeq
    \begin{cases}
        m_a^2-\dfrac{\kappa^2}{m_\gamma^2-m_a^2} & (+) ,\\
        m_\gamma^2+\dfrac{\kappa^2}{m_\gamma^2-m_a^2}& (-) ,
    \end{cases}
\end{align}
where the approximation employed on the right-hand side holds in the weak-coupling limit and as long as the two masses are not degenerate, with $  |\kappa| \ll |m_{\gamma}^2 - m_a^2|$.

Effectively, the background magnetic field mixes the states and generates an effective mass for the photon. The eigenvalues $\lambda_\pm^2$ characterise the dispersion relations of the propagating waves in the medium and should not be literally interpreted as particle masses.
The problem then amounts to computing the free evolution of such states and converting them back to the asymptotic vacuum states.

In the case of multiple axions, the structure is formally equivalent, but now $m_a^2$ is promoted to an $N\times N$  diagonal matrix with entries $m_n^2$ and $\kappa$ is promoted to an $N$-dimensional vector with entries $\kappa_n$ that encodes the axion-photon couplings. The solutions for $\kappa_n\neq 0$ can be found by considering eigenvalues\footnote{See e.g.\ ref.~\cite{deGiorgi:2024elx} for related~discussion.}
\begin{align}
    \label{eq:eigenvalues}&\sum\limits_{n=1}^N\frac{\kappa_n^2}{\lambda^2-m_n^2}=\lambda^2-m_\gamma^2 , \qquad\kappa_n\equiv (c_n g_{a\gamma})\omega B\equiv \omega \beta_n .
\end{align}
The normalized eigenvector corresponding to the eigenvalue $\lambda^2$ reads
\begin{align}
    &u_\lambda=\mathcal{C}_\lambda\begin{pmatrix}
        1\\
        \frac{\kappa_1}{m_1^2-\lambda^2}\\
        \vdots\\
        \frac{\kappa_N}{m_N^2-\lambda^2}
    \end{pmatrix} , \qquad \mathcal{C}_\lambda=\left[1+\sum\limits_{n=1}^N\frac{\kappa_n^2}{(m_n^2-\lambda^2)^2}\right]^{-1/2} .
\end{align}
With this convenient normalization choice, the initial photon state always projects out the $0$-component, which amounts to $\mathcal{C}_0$.

In the regime of weak couplings, the eigenvalues of eq.~\eqref{eq:eigenvalues} can be obtained perturbatively in $\beta_n$. This yields
\begin{align}
    &\lambda_0^2\simeq m_\gamma^2+\sum\limits_{n=1}^N\frac{\kappa_n^2}{m_\gamma^2-m_n^2} , \qquad\lambda_n^2\simeq m_n^2-\frac{\kappa_n^2}{m_\gamma^2-m_n^2} .
\end{align}
This generalizes the single axion result in  eq.~\eqref{eq:single-axion}.
In the limit $\kappa_n\to 0$, which corresponds to vanishing axion-photon mixing, the eigenmode $\lambda_0$ reduces to the standard photon dispersion relation in a medium.
The corresponding normalization factors are
\begin{align}
    &\mathcal{C}_0=1-\frac{1}{2}\sum\limits_{n=1}^N\frac{\kappa_n^2}{(m_\gamma^2-m_n^2)^2}+\mathcal{O}(\kappa^4)  , &\mathcal{C}_n= \left|\frac{\kappa_n}{m_\gamma^2-m_n^2}\right|+\mathcal{O}(\kappa^3) .
\end{align}

In the following sections, we apply this formalism to estimate observable quantities of interest. We first compute the photon disappearance probability that the photon state has oscillated into an orthogonal one after propagation over a distance $L$. Then, we include the regeneration process and estimate the probability of an observable event in an LSW experiment.

\subsection{Axion-photon conversion probability}

We now compute the photon disappearance probability in the presence of multiple axions. We consider an initial state consisting of a pure photon in the interaction basis, which we denote by $\ket{A_0}$. The states $\ket{A_n}$ $(n = 1, \dots, N)$ label the axion components in the interaction basis. Propagation\footnote{For vanishing magnetic field, when $\kappa_n = 0$, the mass basis, the propagation basis and the interaction basis~coincide.} eigenstates of the coupled axion-photon system will be denoted generically by $\ket{A_\lambda}$.

The initial photon state can be decomposed in the propagation basis by means of the unitary transformation $U$ composed of the eigenvectors $u_\lambda$
\begin{equation}
    \ket{A_0} =\sum\limits_\lambda U_{0\lambda} \ket{A_\lambda}=\sum\limits_\lambda \mathcal{C}_\lambda \ket{A_\lambda} .
\end{equation}
Each propagation eigenstate evolves as a plane wave. After propagating over a distance $L$, the state becomes
\begin{equation}
    \ket{A(L,t)}=\sum\limits_\lambda \mathcal{C}_\lambda e^{-i(\omega t-k_\lambda L)}\ket{A_\lambda} .
\end{equation}

Let us first take the photon's effective mass to be non-zero.
The probability that the photon has converted into an orthogonal state is then
\begin{equation}
    P_{\gamma\to X}=1-\left|\braket{A_0|A(L,t)}\right|^2=1-\left|\sum\limits_\lambda \mathcal{C}_\lambda^2 e^{ik_\lambda L}\right|^2=1-\left(\sum\limits_\lambda \mathcal{C}_\lambda^4 +2 \sum\limits_{\lambda'<\lambda} \mathcal{C}_\lambda^2\mathcal{C}_{\lambda'}^2 \cos\left[(k_\lambda-k_{\lambda'}) L\right]\right) .
\end{equation}
The leading contributions are of the order $\kappa^2$. We can extract the photon term with $\mathcal{C}_0\simeq 1+\mathcal{O}(\kappa^2)$ and keep only the leading order terms, giving
\begin{align} \label{eq:dis}
     P_{\gamma\to X}&\simeq 2\sum\limits_{n=1}^N\frac{\kappa_n^2}{(m_\gamma^2-m_n^2)^2} -2\sum\limits_{n=1}^N \mathcal{C}_n^2 \cos\left[(k_n-k_{0}) L\right]  \nonumber\\
      &\nonumber\simeq 2\omega^2\sum\limits_{n=1}^N \frac{\beta_n^2}{(m_\gamma^2-m_n^2)^2} \left(1-\cos\left[\left(\sqrt{\omega^2-m_\gamma^2}-\sqrt{\omega^2-m_n^2}\right) L\right]\right)  \\
     &\simeq \sum\limits_{n=1}^N \frac{4\omega^2\beta_n^2}{(m_\gamma^2-m_n^2)^2}  \sin ^2\left(\frac{m_\gamma^2-m_n^2}{4 \omega }L\right) .
\end{align}
In the last step, we have taken the limit $\omega \gg m_{\gamma}, m_n$ and used $\beta_n \equiv c_n\beta$. This regime is the most relevant one for light axions given the typical energy of lasers in LSW setups $\mathcal{O}(1)$\,eV\@.
For the case of a single axion with mass $m_a$ and $N=1$, in vacuum  $m_\gamma=0$ and relativistic limit $\omega \gg m_a$, the probability yields
\begin{equation}\label{eq:probability_app}
     P_{\gamma\to X}\simeq \frac{4 \beta ^2 \omega ^2}{m_a^4} \sin ^2\left(\frac{L m_a^2}{4 \omega }\right) ,
\end{equation}
in agreement with known results in the literature (see e.g.\ ref.~\cite{Arias:2010bh}).

We can generalize the formalism to account for propagation in a refractive and absorptive medium. In particular, this requires appropriate care when evaluating the modulus squared.
In an absorptive medium, the effective photon mass acquires an imaginary component, which acts as a damping term as photons get absorbed by the medium
\begin{equation}
    m_\gamma^2=m_{\gamma R}^2-i\Gamma \omega ,
\end{equation}
where $m_{\gamma R}$ is the real part of the effective photon mass induced by medium refraction, and
$\Gamma$ denotes the photon absorption coefficient of the medium.\footnote{The quantity $\Gamma^{-1}$ corresponds to the photon attenuation length, such that in the absence of axion mixing the photon intensity decays as~$e^{-\Gamma z}$.}

As we will also see below, in axion searches, the refractive effect can be useful. By tuning the effective photon mass (e.g.\ if the medium is a gas, adjusting its pressure), it is possible to match the axion mass, enhancing coherence and thus enhancing conversion probability.
The photon disappearance probability in the high-energy limit becomes
\begin{align} \label{eq:disabs}
   P_{\gamma \to X}\simeq \left(1-e^{-\Gamma L}\right)+ \sum\limits_{n=1}^N \dfrac{2 \beta_n^2 e^{-\frac{\Gamma  L}{2}} }{\left(\Gamma ^2+4 q_n^2\right)^2} \left\{ 4 \Gamma  q_n \sin \left(q_nL\right)+\left(\Gamma ^2-4 q_n^2\right)\left[\cos \left(q_n L\right)-e^{-\frac{\Gamma  L}{2}}\right] \right\}
\end{align}
where
\begin{equation}
    q_n\equiv \frac{|m_n^2-m_{\gamma R}^2|}{2\omega} .
\end{equation}
In the limit $\Gamma \to 0$ eq.~\eqref{eq:disabs} reduces to eq.~\eqref{eq:dis}.
The term independent of $\beta_n$ takes into account the absorption of the medium, which allows for photon disappearance even in the absence of axions.
Similarly, the axion-photon conversion from a pure $a_n$ initial state in the high-energy limit reads
\begin{align}\label{eq:P_medium}
     P_{a_n\to \gamma}&\simeq\left(\frac{\beta_n}{2}\right)^2 \dfrac{1-2 e^{\frac{-\Gamma  L}{2}}  \cos (q_n L)+ e^{-\Gamma L}}{q_n^2+\Gamma ^2/4},
\end{align}
which again agrees with known results in the literature~\cite{CAST:2008ixs}.

\subsection{Axion production and photon regeneration}
Let us now focus on an LSW experiment setup, as schematically depicted in figure~\ref{fig:LSW-schematic}. As we will show, in the presence of multiple axions, the wall plays a non-trivial role. Although photons are absorbed by the wall, axion propagation and oscillations among ``sterile'' axion states continue across the wall region even in the absence of a magnetic field. Thus, we include the wall's thickness as a parameter in the evolution.

We denote by $T(x_i,x_f)$ the spatial evolution operator from position $x_i$ to $x_f$.
We choose coordinates such that the laser source is located at $x_0=0$, the first magnetic field region of length $L_1$ ends at $x_1$, the wall extends from $x_1$ to $x_w$ over a distance $L_w$, and the second magnetic field region ends after a distance $L_2$ at $x_2$ where the detector is assumed to be located.
Since the global factor $\exp(-i\omega t)$ is common to all components, it is factorized and omitted in the following.
The evolution proceeds through three distinct regions:
\begin{enumerate}
    \item \textbf{Magnetic field region 1} ($0 \to x_1$, length $L_1$):
    the initial photon state $\ket{A_0}$ evolves into a superposition of propagation eigenstates
    \begin{align}
  &\ket{A(x_1)}=T(0,x_1)\ket{A_0}= \sum\limits_\alpha  U_{0\alpha} T_{01,\alpha}\ket{A_\alpha} , \qquad T_{01,\alpha}=\exp{(ik_\alpha L_1)} .
    \end{align}
    \item \textbf{Wall region} ($x_1 \to x_w$, length $L_w$): the wall acts effectively as a measurement device, absorbing photons and projecting out the photon component $A_0$.
The state after is
        \begin{align}
  &\ket{A(x_1)}= \sum\limits_\alpha  U_{0\alpha} T_{01,\alpha}\ket{A_\alpha} \to\ket{A(x_1)}_R= R\sum\limits_\alpha \sum\limits_{n=1} U_{0\alpha}U_{n\alpha}^\star T_{01,\alpha}\ket{A_n} ,
    \end{align}
    where $R$ accounts for the new normalisation of the state. Then, the axion components propagate freely across and fields evolve in the interaction basis from $x_1$ to $x_w$
    \begin{align}
        &\ket{A(x_w)}=T(x_1,x_w)\ket{A(L_1)}_R=R\sum\limits_\alpha \sum\limits_{n=1} U_{0\alpha}U_{n\alpha}^\star T_{01,\alpha}T_{1w,n}\ket{A_n} ,\\ \nonumber &T_{1w,n}=\exp{(ik_n L_w)} .
    \end{align}
    \item \textbf{Magnetic field region 2} ($x_w \to x_2$, length $L_2$): in the second magnetic field region the axions can reconvert into photons analogously,
        \begin{align}
        &\ket{A(x_2)}=T(x_w,x_2)\ket{A(x_w)}_R=R\sum\limits_{\alpha\beta} \sum\limits_{n=1} U_{0\alpha}U_{n\alpha}^\star U_{n\beta}T_{01,\alpha}T_{1w,n}T_{w2,\beta}\ket{A_\beta} , \\
        &\nonumber T_{w2,\beta}=\exp{(ik_\beta L_2)} .
    \end{align}
    We included a coefficient $R$ to account for the potential normalization of the new state.
\end{enumerate}

The resulting photon regeneration probability is obtained by projecting back onto the photon state $A_0$, obtaining
    \begin{equation}
        P_{\gamma\to X\to \gamma}=|\braket{A_0|A(x_2)}|^2=|R|^2\left|\sum\limits_{\alpha\beta} \sum\limits_{n=1} (U_{0\alpha}U_{n\alpha}^\star U_{n\beta}U_{0\beta}^\star) T_{01,\alpha}T_{1w,n}T_{w2,\beta}\right|^2 .
    \end{equation}
Since we are already taking into account the loss of signal due to photon absorption by the wall, we set $R=1$.

For simplicity, we now work in vacuum and neglect absorptive effects, setting $m_\gamma=0$.
Considering the structure of computations, a central quantity is
\begin{align}
    \sum\limits_{\alpha}U_{0\alpha}U_{n\alpha}^\star T_{01,\alpha}&=\kappa_n\sum\limits_{\alpha} \frac{\mathcal{C}_\alpha^2}{m_n^2-\lambda_\alpha^2} \exp{(ik_\alpha L_1)} \simeq\frac{\kappa_n}{m_n^2}\left(e^{iL_1\omega}-e^{iL_1\sqrt{\omega^2-m_n^2}}\right) +\mathcal{O}(\kappa^2)\\
    &\nonumber\simeq \frac{2 i \kappa_n}{m_n^2}\sin\left(\frac{m_n^2L_1}{4\omega}\right) e^{iL_1[\omega-m_n^2/(4\omega)]} ,
\end{align}
where in the last step we took the limit $\omega \gg m_n$.

Substituting the above into the probability equation,  and considering the same limits, we obtain
\begin{align}
    \label{eq:probability-formula}
        P_{\gamma\to X\to \gamma}&\simeq\left|\sum\limits_{n=1} \frac{\kappa_n^2}{m_n^4}\left(e^{iL_1\omega}-e^{iL_1k_n}\right)\left(e^{iL_2\omega}-e^{iL_2k_n}\right)\exp{(ik_n L_w)}\right|^2  \\&\nonumber\simeq\left|\sum\limits_{n=1} \frac{4\kappa_n^2}{m_n^4}\sin\left(\frac{m_n^2L_1}{4\omega}\right)\sin\left(\frac{m_n^2L_2}{4\omega}\right)e^{-im_n^2(L_1+2L_w+L_2)/(4\omega)]}\right|^2  \\
        &\nonumber=\left[\sum\limits_{n=1} P_{\gamma\to n}(L_1)P_{\gamma\to n}(L_2)+2\sum\limits_{k<n} \Pi_{1,k}^\star\Pi_{2,k}^\star\Pi_{1,n}\Pi_{2,n}\cos\left(\frac{m_n^2-m_k^2}{2\omega}\left(\frac{L_1+L_2}{2}+L_w\right)\right)\right] ,
    \end{align}
    where
    \begin{align}
        &\label{eq:Pi-function}\Pi_{a,n}\equiv \frac{2\omega \beta_n}{m_n^2}\sin\left(\frac{m_n^2L_a}{4\omega}\right) , \qquad|\Pi_{a,n}|^2=P_{\gamma\to n}(L_a) .
    \end{align}
In eq.~\eqref{eq:probability-formula}, the first term of the expanded sum corresponds to the incoherent sum of the individual axion contributions. Diagrammatically, this is equivalent to considering that photons oscillate into specific axion mass eigenstates, which subsequently reconvert into photons. The second term encodes interference between different axion mass eigenstates.
This interference arises because the photon oscillates into a superposition of axion states rather than a single eigenstate.
The axion states can oscillate from one to another.
Since this is an effect also present in the vacuum and axion propagation continues inside the wall, the wall thickness contributes to the interference phase and thus affects the final probability. This phenomenon has no single particle analogue and therefore can be used to distinguish between single and multiple axion scenarios. We will discuss this in section~\ref{sec:fourier}. We note that if the squared masses of multiple axions are approximately degenerate or both very light compared to $\omega/L_{1,2,w}$, then no relative phases are accumulated during propagation through the wall, and the interference terms reduce to a coherent enhancement equivalent to that of a single effective axion.

Let us consider the limits of the probability formula. It suffices to consider $\Pi_n(L)$ defined in eq.~\eqref{eq:Pi-function}. Let us define
\begin{equation}
    r_n\equiv\frac{m_n^2L}{4\omega}\simeq 1.3 \left(\frac{m_n}{10^{-4}~\text{eV}}\right)^2\left(\frac{\text{eV}}{\omega}\right)\left(\frac{L}{100~\text{m}}\right)\,.
\end{equation}
In the limit $r_n\gg 1$, the sine function becomes highly oscillatory over the experimental baseline and the probability becomes suppressed by the squared mass of the fields.
In the limit $r_n\ll 1$  one has
\begin{align}
    &\Pi_n(L)\simeq \beta_n\frac{L}{2} , \qquad P_{\gamma\to a_n}^2 \simeq \beta_n^4\frac{L^4}{16} .
\end{align}
Therefore, as $m_n \to 0$, the probability becomes independent of the axion mass and grows with the size of the active magnetic region. Hence, for non-hierarchical couplings $c_n$, the probability and total signal are dominated by the lightest modes satisfying $r_n \ll 1$.

To illustrate the role of interference effects let us consider a regime in which $r_n \ll 1$, with the wall thickness $L_w$ sufficiently large such that the relative phases $(m_n^2 - m_k^2)L_w/(2\omega)$ can become appreciable.\footnote{In practice, for a realistic setup increasing the wall thickness to this extreme is not necessarily experimentally~advantageous.}
In this case, the probability can be approximated as
    \begin{align}
    \label{eq:probability-formula-approx}
        P_{\gamma\to X\to \gamma}&\simeq\left(\beta^4\frac{L^4}{16}\right)\left[\sum\limits_{n,\text{light}} c_n^4+2\sum\limits_{k<n,\text{light}} c_k^2 c_n^2\cos\left(\frac{m_n^2-m_k^2}{2\omega}L_w\right)\right] .
    \end{align}
The first part is a sum of individual probability contributions and therefore ``incoherent''. Relative to the single axion scenario, it is enhanced by a factor of $N_{\text{light}}$ when all couplings are comparable. The second term encodes interference effects arising from the superposition of different axion contributions.

When the phase differences in eq.~\eqref{eq:probability-formula-approx} are small, the contributions add coherently, and the light fields contribute with an $N_\text{light}^2$ enhancement
    \begin{align}
    \label{eq:probability-formula-approx-coherent}
        P_{\gamma\to X\to \gamma}&\simeq\left(\beta^4\frac{L^4}{16}\right)\left(\sum\limits_{n,\text{light}} c_n^2\right)^2 \propto N_\text{light}^2 .
    \end{align}
The results are in agreement with the expectations from eq.~\eqref{eq:estimation-sigma}. While this coherent regime maximally enhances the signal, the information about the individual axion masses is hidden, rendering the experiment sensitive only to the squared sum of the squared coefficients of the coupling coefficients $c_n$. Information about the axion masses is accessible once the wall thickness is sufficiently large for the interference phases to become appreciable. We discuss practical strategies for exploiting this effect in section~\ref{sec:fourier}.

Finally, we emphasize that interference effects can also be destructive. In extreme cases, they can lead to an entirely vanishing regenerated photon conversion probability and signal. As a result, the presence of multiple axions can reduce, rather than enhance, the sensitivity of a given experiment (a similar discussion will be
presented in the context of dark photons in an upcoming work~\cite{Monath}).

\subsection{Comparison with neutrino oscillations}
\label{ssec:neutrinos}

The formalism we developed for axion-photon oscillations is closely analogous to the known neutrino flavour oscillations~\cite{Pontecorvo:1957qd,Maki:1962mu,Super-Kamiokande:1998kpq,Esteban:2024eli}, including the neutrino matter-enhanced propagation, i.e.\ the Mikheyev-Smirnov-Wolfenstein (MSW) effect~\cite{Mikheyev:1985zog,Wolfenstein:1977ue}.
Here, we comment on the key similarities and qualitative differences relevant for oscillations of multiple axions.

Consider an electron flavour neutrino produced as a weak eigenstate $|\nu_e\rangle$, with flavour basis states $|\nu_e\rangle$, $|\nu_\mu\rangle$, $|\nu_\tau\rangle$.
Its propagation and evolution in time is governed by superposition of mass eigenstates $|\nu_1\rangle$, $|\nu_2\rangle$, $|\nu_3\rangle$
\begin{equation}
|\nu(t)\rangle = \sum_{i=1}^3 U_{ei}^* e^{-iE_it} |\nu_i\rangle,
\qquad\text{with}\quad
|\nu_e\rangle = \sum_{i=1}^3 U_{ei} |\nu_i\rangle\,,
\label{eq:nu_state_evolution}
\end{equation}
where $U$ is the unitary $3\times3$ Pontecorvo-Maki-Nakagawa-Sakata (PMNS) matrix and $E_i$ is the energy of $i$-th mass  eigenstate.
The probability of detecting another flavour follows from the overlap with the corresponding weak eigenstate.
For $\nu_e\to\nu_\mu$ in the 2-flavour vacuum limit\footnote{We consider that neutrinos are always~relativistic.}
\begin{equation} \label{eq:nuprob}
P_{\nu_e\to\nu_\mu}
= \bigl|\langle\nu_\mu|\nu(t)\rangle\bigr|^2
= \bigl|\sum_{i} U_{\mu i}  U_{ei}^*  e^{-iE_i t}\bigr|^2
\simeq \sin^2 2\theta  \sin^2  \left(\frac{\Delta m^2 L}{4E}\right)\,,
\end{equation}
where for relativistic neutrinos time evolution is approximately tracked by the distance traveled $t\simeq L$ and, for fixed momentum $p\simeq E$, $E_i\simeq E+m_i^2/(2E)$,
$\sin^2 2\theta$ describes the flavour mixing, $\Delta m^2$ denotes the mass squared difference between two neutrino mass eigenstates, $L$ is the propagation length and $E$ is the neutrino energy. Eq.~\eqref{eq:nuprob} is structurally identical to the single axion conversion probability of eq.~\eqref{eq:probability_app} if we consider the phase $(\Delta m^2 L)/(4E)$ replaced by $(m_\gamma^2 - m_n^2)L/(4\omega)$.
The generalization to multiple axions then follows the multiple-flavour neutrino case. There, the initial photon state $|A_0 \rangle$ projects onto several propagation eigenstates $|A_{\lambda}\rangle$ with coefficients determined by the unitary mixing matrix $U$ and each acquiring a phase $e^{-ik_\lambda L}$. The observable conversion rate then follows from projecting back onto the photon state.

In matter, forward scattering of $\nu_e$ on electrons with density $n_e$ induces an effective potential $A = \sqrt{2} G_F n_e$, with $G_F$ being the Fermi weak coupling constant.
In the 2-flavour case, this is described by the Hamiltonian
\begin{equation}
H=\frac{\Delta m^2}{4E}
\begin{pmatrix}
-\cos2\theta & \sin2\theta \\
\sin2\theta & \cos2\theta
\end{pmatrix}
+ A
\begin{pmatrix}
1 & 0 \\
0 & 0
\end{pmatrix},
\end{equation}
leading to resonant conversion when $A \simeq \Delta m^2\cos2\theta/(2E)$.
In the axion case, the external magnetic field $B$ provides the off-diagonal mixing $\kappa_n = g_{a_n\gamma} \omega B$ as in eq.~\eqref{eq:alpmixing}. When the effective photon mass is modified to $m_\gamma^2 \neq 0$, such as in a buffer gas environment as used in section~\ref{sec:helioscopes}, this is analogous to the MSW potential affecting neutrino propagation in matter. As we discuss, variation of the gas pressure in medium therefore modifies the effective axion-photon mixing angle and can generate resonant behaviour in close parallel to matter neutrino oscillation enhancements.

Despite similarities, there are important qualitative differences between axion-photon and neutrino mixing oscillations.
First, neutrinos undergo flavour mixing oscillations in vacuum, while axion-photon mixing is induced by the external magnetic field. This vanishes when $B=0$, but it can be controlled in an experimental setup.\footnote{In contrast to the situation in astrophysical axion~searches.}
Second, for neutrinos, the matter potential modifies propagation through the MSW effect continuously while preserving unitarity and does not remove any flavor components. This is in contrast to the potential photon absorption in a medium.
Moreover, for axion-photon mixing, such as in LSW experiments, the wall absorbs the photon component and projects the system exclusively onto the axion subspace.
This effectively signifies a non-unitary projection of state $|A_0\rangle \rightarrow 0$ rather than an MSW-like modification of the Hamiltonian.
After traversing an opaque wall of thickness $L_w$, the state for $N$ axions is
$|\psi(L_w)\rangle
= \sum_{n=1}^N c_n e^{-ik_n L_w} |a_n\rangle$
with $k_n\simeq\sqrt{\omega^2-m_n^2}$ and with no surviving photon component.
This projection has no analogue in standard neutrino oscillations and introduces relative phases between different axion modes that can be controlled.

\section{Benchmark models}
\label{sec:models}

In this section, we introduce several representative models that we use as benchmarks to illustrate the qualitative behaviour of multiple axions in laboratory experiments. Since the effects discussed in this work become most pronounced when multiple fields are present, we focus on scenarios that have a large number of axions. Specifically, we consider an axiverse-inspired scenario and two scenarios motivated by extra-dimensional constructions.

For the string theory axiverse-inspired benchmark~\cite{Svrcek:2006yi,Arvanitaki:2009fg}, which we call ``Stringy Axions'', we adopt for reference a log-normal distribution of axion masses with a uniform sampling of the action motivated by non-perturbative effects in string theory compactifications.  The axion masses and decay constants $f_n$ are parametrized as
\begin{align}
    &m_n = \mathcal{O}(1)\times \left(\frac{\mu^2}{f_n}\right) e^{-S/2}
 , &f_n = \mathcal{O}(1)\times \frac{\Lambda}{S} ,
\end{align}
where $S$ is the string instanton action.  For concreteness, we take
\begin{align}
  &\label{eq:distributions}\mu=\Bar{M}_P , &&\Lambda=10^{11}~\text{GeV} ,   &&S\in[150,250] ,
\end{align}
with $\bar{M}_P \simeq 2.4\times10^{18}~\mathrm{GeV}$ being the reduced Planck mass. We stress that the parameter choices adopted here are intended to be illustrative and motivated by classes of string-inspired models, serving primarily to highlight the potential phenomenological impact of such constructions in laboratory settings. We sample $S$ from a uniform distribution within the interval specified above, although larger values of $S$ are   possible~\cite{Svrcek:2006yi,Arvanitaki:2009fg}. We randomly generate $\mathcal{O}(1)$ coefficients entering $m_n$ and $f_n$ employing a log-normal distribution whose exponent is centred at zero with width $0.1$, corresponding to variations within approximately a factor of two.

\begin{table}
    \centering
\resizebox*{1\textwidth}{!}{    \begin{tabular}{r|ccc}
    \toprule
        Model  & Stringy Axions & KK Maxions  & KK ALPs \\
        \midrule
        Mass ($m_n$) &$\sim \mu^2  e^{-S/2}/f_n$ &$\left(n-\frac{1}{2}\right)\mu_1$& $\sqrt{m_0^2+(n\mu_1)^2}$  \\
        Decay constant ($f_n$) &$\sim\Lambda/S$&$f_a\times\frac{\sqrt{2}}{2n-1}\left(\frac{\pi m_a}{\mu_1}\right)^2$& $f_a=\text{const.}$ \\
        \midrule
        Input values (LSW): & $\begin{cases}
            \mu=\Bar{M}_P ,\\
            \Lambda=10^{11}~\text{GeV}  \\
            S\in[150,250]
        \end{cases}$ &$\begin{cases}
            \mu_1=5\times10^{-5}~\text{eV}  \\
            m_a=\mu_1/\pi  \\
            f_a =10^{8}~\text{GeV}
        \end{cases}$ & $\begin{cases}
             \mu_1=10^{-6}~\text{eV}  \\
            m_0=10^{-4}~\text{eV}  \\
            f_a =10^{8}~\text{GeV}
        \end{cases}$\\
                \midrule
        Input values (helioscopes): & $\begin{cases}
            \mu=\Bar{M}_P  \\
            \Lambda=10^{11}~\text{GeV}  \\
            S\in[150,250]
        \end{cases}$ &$\begin{cases}
            \mu_1=5\times10^{-2}~\text{eV}  \\
            m_a=\mu_1/\pi  \\
            f_a =10^{8.5}~\text{GeV}
        \end{cases}$ & $\begin{cases}
             \mu_1=10^{-3}~\text{eV}  \\
            m_0=10^{-1}~\text{eV}  \\
            f_a =10^{8}~\text{GeV}
        \end{cases}$\\

        \bottomrule
    \end{tabular}
}\relax    \caption{Summary of the benchmark models adopted in this work. The listed values specify the input parameters used for the LSW and helioscope analyses, respectively, and are chosen to highlight characteristic features of multiple axion~signals.\label{table-parameters}}
\end{table}

We next consider two benchmark scenarios inspired by extra-dimensional axions~\cite{Dienes:1999gw,Dienes:2011ja,Dienes:2011sa,deGiorgi:2024elx}, in which the axion fields arise as Kaluza-Klein (KK) modes. We assume KK contributions to the masses are equally spaced by the mass of the lightest mode, here denoted by $\mu_1$, so that $\mu_n= n\mu_1$. We consider the effective four-dimensional Lagrangian
\begin{align}
 &\mathcal{L}_{\rm eff} =\sum\limits_{n=1}^N\left(\frac{1}{2}\partial^\mu a_n \partial_\mu a_n-\frac{1}{2}(m_0^2+\mu_n^2)  a_n^2 \right)+ \Lambda_I^4 \cos\left(\frac{1}{f_a}\sum\limits_{n=1}^N \psi_n a_n\right) ,
\end{align}
where $\psi_n$ are model-dependent coefficients and $\Lambda_I$ denotes a non-perturbative instanton scale.
The expansion of the cosine potential generates an additional mass term for the linear combination of axions
\begin{align}
    &\mathcal{L}_{\rm eff}\supset -\frac{1}{2}m_a^2 \left(\sum\limits_{n=1}^N \psi_n a_n\right)^2 \,, &m_a^2\equiv \frac{\Lambda_I^4}{f_a^2}\,.
\end{align}
Upon diagonalising the full mass matrix, including potential contribution from $\Lambda_I$, this setup can be mapped onto the effective Lagrangian of eq.~\eqref{eq:Lagr}.
We focus on the two cases ($m_0=0,\Lambda_I\neq 0$; ``KK  Maxion'') and ($m_0\neq 0,\Lambda_I= 0$; ``KK ALP''). In the limit $\mu_1\ll m_0$, mass degeneracies among the modes appear. Similarly, if $\mu_1\ll m_a$, the modes maximally mix in the mass matrix.
Analytic expressions for such coefficients for certain extra-dimensional models for $m_0=0$ have been considered (e.g.\ ref.~\cite{deGiorgi:2024elx}), and can be generalized for $m_0\neq 0$.

\begin{figure}
    \centering
    \includegraphics[width=\linewidth]{ 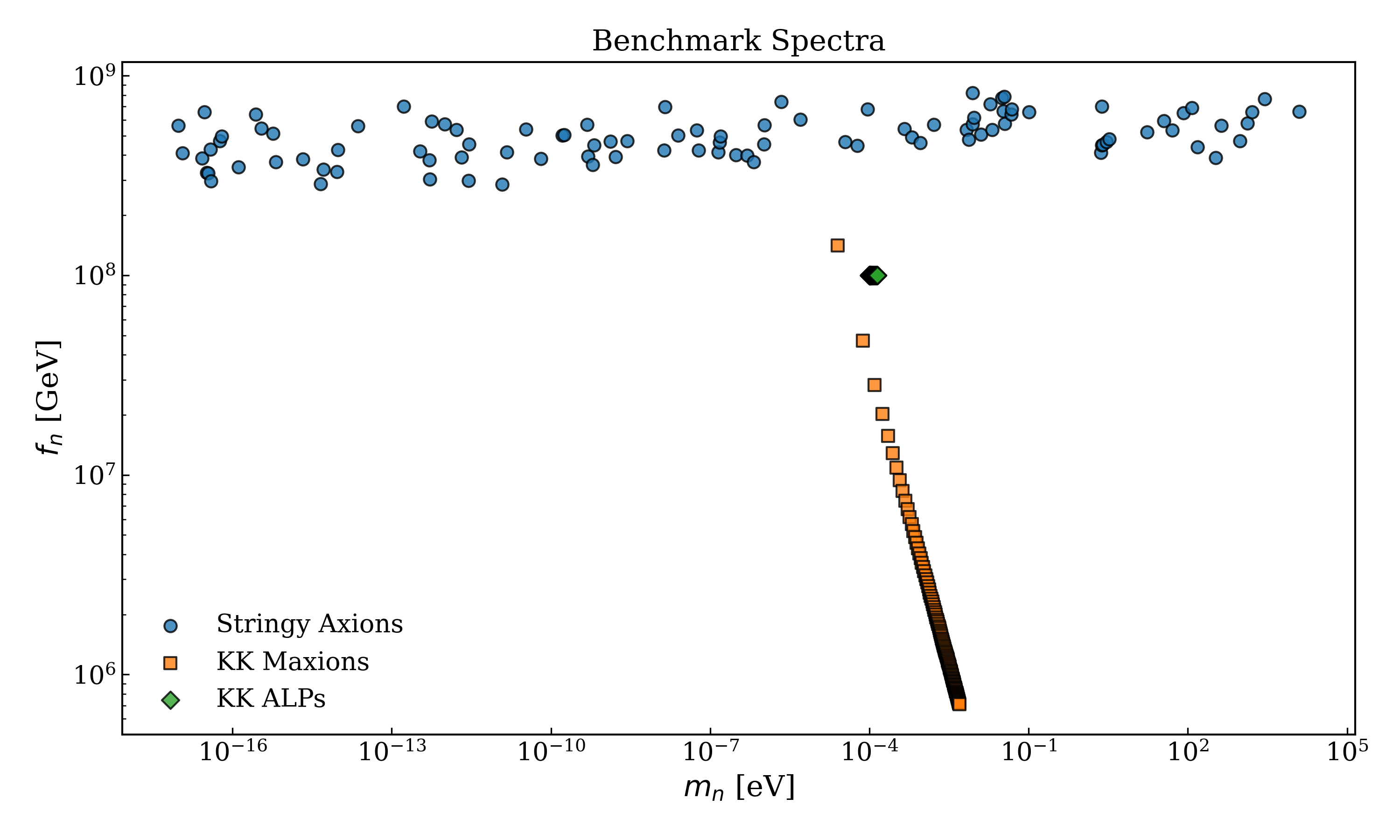}
    \caption{Representative spectra of axion masses and associated scales in the benchmark scenarios considered in this work, including an axiverse-inspired log-normal distribution (``Stringy Axions'', blue circles), a KK-Maxion-type spectrum (``KK Maxions'', orange squares)  and a KK-type spectrum with a tree-level mass (``KK ALPs'', green diamonds), considering $N = 100$ axions. See  table~\ref{table-parameters} for additional LSW parameter~details.\label{fig:parameters-example}}
\end{figure}

In table~\ref{table-parameters} we summarize the parameter choices for benchmark models we adopt.  Representative examples of the resulting mass spectra and associated scales are shown in figure~\ref{fig:parameters-example} for $N=100$ axions.  The precise relation between $f_n$ and axion-photon coupling $g_{a_n\gamma}$ depends on the ultraviolet-complete model. In the following, for illustration, we will employ the relation
\begin{equation}
    g_{a_n\gamma}\equiv c_n g_{a\gamma}\equiv  \frac{\alpha}{2\pi}\frac{1.92}{f_n} ,
\end{equation}
which corresponds to the irreducible contribution to the axion-photon coupling induced by pion mixing in the canonical QCD axion scenario (cf., e.g.~ref.~\cite{GrillidiCortona:2015jxo}). While this relation is not necessarily universal for generic axion-like particles, it provides a representative normalization for an estimate. That said, in string-axiverse scenarios, typically not all axions will inherit $\mathcal{O}(1)$ couplings to photons~\cite{Gendler:2023kjt}. Furthermore, under a number of assumptions, ALPs above the QCD line are restricted if they stem from a simple ultraviolet Grand Unification Theory in four   dimensions~\cite{Agrawal:2022lsp}.

\section{Light shining through wall experiments}
\label{sec:LSW}

In this section, we apply the general probability formula derived in eq.~\eqref{eq:probability-formula} to LSW experiments. In this class of experiments, axions are produced and detected entirely within a controlled laboratory environment. This is in contrast to helioscope and haloscope experiments, which rely on axions originating from the Sun and from DM, respectively. The ability to control both production and detection conditions provides a distinct advantage, and additional handles can be exploited to extract detailed information about the properties of the exchanged particles, including the presence of multiple axion states. On the other hand, in LSW experiments, both production and regeneration rates are suppressed by the small axion-photon coupling.

In the following, we adopt the ALPS~II experiment~\cite{Ortiz:2020tgs}\footnote{Notable past experiments include BFRT~\cite{Cameron:1993mr}, GammeV~\cite{Steffen:2009sc}, ALPS~\cite{ALPS:2009des}, OSQAR~\cite{OSQAR:2015qdv}.} that is expected to achieve significant sensitivity (see~\cite{ALPSII:2025eri} for first science results) as a representative benchmark.
 We consider the experimental configuration of ALPS~II\footnote{We thank A.~Lindner for additional helpful information, in particular on the wall~thickness.}~\cite{Ortiz:2020tgs,Kozlowski:2024jzm} employing production and regeneration cavities~(PC and RC, respectively) of equal length $L=106$~m, a high-power laser with a wavelength of $\lambda=1064$~nm  with a design objective power of $P_\text{PC}=150$~kW and a magnetic field of $B=5.3$~T.

\subsection{Signal strength in benchmark models}

For a single very light axion, the photon regeneration probability is
\begin{equation}
    P_{\gamma \to a\to\gamma}\simeq \frac{1}{16}(\beta L)^4,
\end{equation}
which we use as a reference for comparison.
The expected number of photon events $n_{\rm ev}$ for a measurement of duration $\Delta t$ is given by~\cite{Ortiz:2020tgs},
\begin{align}
\label{eq:signal}
    n_{\rm ev}&= n_\gamma P_{\gamma \to a\to\gamma} \simeq \left(10^{33}  P_{\gamma \to a\to\gamma}\right) \left(\frac{\Delta t}{40~\text{hours}}\right)\left(\frac{1.16~\text{eV}}{\omega}\right)\left(\frac{P_\text{PC}}{150~\text{kW}}\right)\left(\frac{\beta_\text{RC}}{10^4}\right)\left(\frac{\eta}{0.9}\right) ,\notag\\
    & \simeq1 \left(\frac{\Delta t}{40~\text{hours}}\right)\left(\frac{1.16~\text{eV}}{\omega}\right)\left(\frac{P_\text{PC}}{150~\text{kW}}\right)\left(\frac{\beta_\text{RC}}{10^4}\right)\left(\frac{\eta}{0.9}\right)\left(\frac{g_{a\gamma}~\text{GeV}}{2\times 10^{-11}}\right)^4\left(\frac{B}{5.3~\text{T}}\right)^4 \left(\frac{L}{106~\text{m}}\right)^4 \,,
\end{align}
where $\eta$ denotes the coupling efficiency between the PC and RC and $\beta_\text{RC}$ is the
power build-up factor of RC.

\begin{figure}
    \centering
    \includegraphics[width=0.9\linewidth]{ 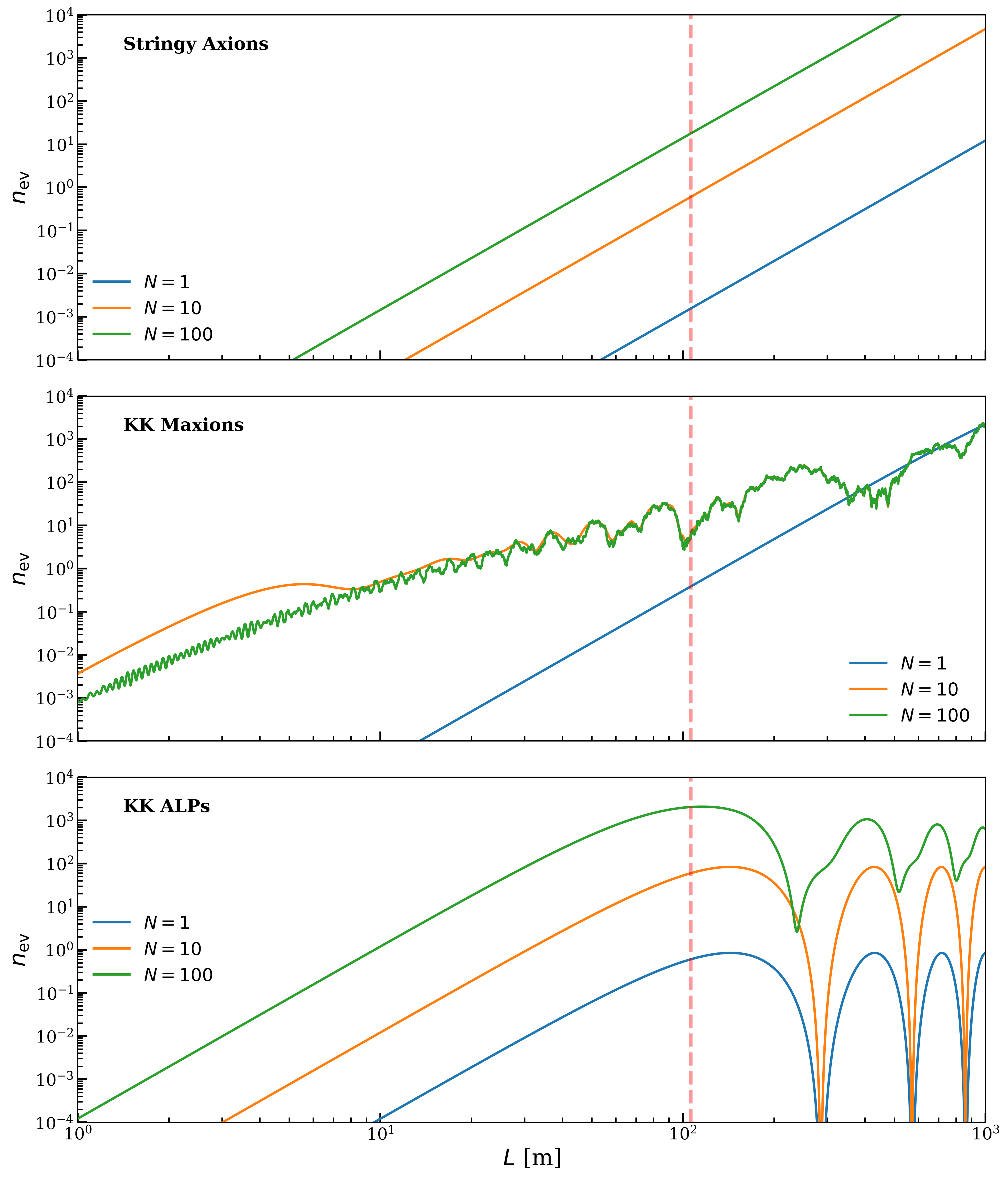}
    \caption{Expected number of regenerated photon events $n_{\rm ev}$ for an observation time of $\Delta t=40$~hr as a function of the length of the symmetric magnetic field region $L$ for the benchmark models introduced in section~\ref{sec:models}, assuming the ALPS~II experimental configuration. The nominal ALPS~II design baseline $L=106$~m is highlighted by a red dashed~line.\label{fig:signals-models}}
\end{figure}

In figure~\ref{fig:signals-models}, considering the ALPS~II configuration, the expected number of resulting regenerated photon events is shown for the benchmark models introduced in section~\ref{sec:models} as a function of the magnetic field region length $L_1 = L_2 = L$. We showcase for reference lengths of $L$ up to 1~km. While ambitious, technical feasibility has been considered in e.g.~\cite{Arias:2010bh,Hoof:2024gfk}. We show results for scenarios with $N = 1, 10, 100$ axion states included in the calculation.\footnote{That is, we cut off other axion contributions. The lighter modes are added first as $N$ is~increased.} The stringy axion, KK Maxion and KK ALP benchmarks correspond to the top, middle and bottom panels, respectively.

The stringy axion benchmark is shown in the top panel of figure~\ref{fig:signals-models}. Since all couplings are considered to be of the same order of magnitude, the enhancement essentially only depends on the number of axion species at each mass scale. For $N=1$, the signal exhibits the characteristic $\propto L^4$ scaling of a single light axion. As additional light modes are included, coherence among the states leads to a clear $\propto N^2$ enhancement, visible when comparing the $N= 1$ and $N = 10$ cases. Upon including heavier modes, coherence is gradually lost, and the enhancement transitions to linear scaling with $\propto N$, as illustrated by the $N = 100$ curve.

The middle panel of figure~\ref{fig:signals-models} shows the KK Maxion scenario, where the interplay between mass splittings and couplings leads to a more intricate behaviour.
For $N=1$ the situation is analogous to the stringy axion scenario. For small $L$, multiple modes remain effectively light but with larger coupling, and coherently enhance the signal, resulting initially in an $\propto N^2$ enhancement
followed by a quartic enhancement due to smaller $f_n$. At larger $L$, the probability contribution of heavier modes is suppressed by the eighth power of their masses $m_n$. Thus, the probability reduces to the $N=1$ case modulated by the effect of the oscillatory behaviour of the other axion fields. For $N=100$, the conversion probability is suppressed even for small values of $L$. This is due to the structure of the maxions spectrum since as modes get more massive  their coupling grows. Their contribution to the signal becomes sizable but rapidly oscillating, thus strengthening destructive interference and suppressing the signal. For the chosen parameters, we see this suppression sets in as soon as the factors $ 4\omega/L$ and $m_n^2$ in eq.~\eqref{eq:probability-formula} become of the same order. Thus, at $L\sim 7~\text{m}$, all but the lowest 10 modes ($m_{n \leq  10} < 52.5\times10^{-5}\text{ eV}$) contribute in the form of rapid oscillations. At $L \sim  300$~m, only the lowest mode in the spectrum ($m_{n=1} < 7.5\times10^{-5}\text{ eV}$) remains unsuppressed. This behaviour arises from decoherence effects, which for certain mass configurations partially cancel the coherent contributions and reduce the effective linear growth of the probability. This illustrates that increasing the number of axion fields does not generically enhance the signal. For larger values of $L$, the behaviour approaches the previously discussed regime, characterized by increased oscillations.

The bottom panel of figure~\ref{fig:signals-models}, focusing on the KK ALP scenario, clearly illustrates the role of mass splittings in determining whether the signal exhibits an $N$ or $N^2$ enhancement. For $N=1$ and $N=10$, the mass splitting between the lightest and heaviest modes, $\Delta m^2 \sim N^2 \mu_1^2 \sim N^2 \times 10^{-12} ~\mathrm{eV}^2 \sim 10^{-10}~ \mathrm{eV}^2$, is sufficiently small to maintain coherence among the axion fields over the entire range of lengths shown, resulting in the expected $N^2$ scaling. By contrast, for $N=100$ the splitting increases to $\Delta m^2 \sim 10^{-8}~ \mathrm{eV}^2$, such that, for the chosen benchmark parameters, $N^2$ coherence is preserved only for $L \lesssim 100 \,\mathrm{m}$. This behaviour is evident from the comparison of the left and right regions of the plot, where the ratio of the $N=100$ to $N=10$ signals transitions from $\sim N^2 = 10^2$ to $\sim N = 10$. For the present benchmark choice, this change in scaling coincides with the onset of oscillatory behaviour associated with the value of $m_0$.

\enlargethispage*{\baselineskip}\relax
Beyond coherence effects, subleading modes can also contribute sufficiently to induce apparent modulation effects. This is visible in the KK Maxion scenario, where the $N=100$ case closely follows the $N=10$ signal, up to small, rapidly oscillating contributions.

\subsection{Fourier analysis for parameter determination}
\label{sec:fourier}

A possible approach to gain access to more information on the mass spectrum and the coupling structure is by adapting and varying experimental configurations.
Since in LSW experiments laser energies are typically constrained, the relevant experimental parameters to consider for variation are the lengths of the active magnetic regions and the thickness of the wall, as well as, potentially, the refractive properties of the conversion region. By performing measurements over a range of such parameters, it is in principle possible to reconstruct features of the underlying axion spectrum by employing techniques such as Fourier analysis or a fit to model expectations. In this section, we illustrate the potential to obtain more information by using Fourier analysis.  We describe the resulting spectral structure in an idealized setting and comment on some practical limitations.

For illustration, we consider a signal arising from a known axion mass spectrum. In an ideal case, experimentally, we could vary the length of the active magnetic field regions and of the wall's thickness continuously and reconstruct the Fourier transform $\mathcal{F}$ of the signal $f(\ell)$, with respect to the Fourier mode $k$, given by
\begin{equation}
    \mathcal{F}[f(\ell)](k)\equiv\hat{f}(k)=\frac{1}{\sqrt{2\pi}}\int\limits_{-\infty}^\infty d\ell e^{-ik \ell}f(\ell).
\end{equation}
Since the photon regeneration probability in eq.~\eqref{eq:probability-formula} is composed of sine and cosine functions, its Fourier transform consists of a discrete set of Dirac delta peaks.

Separating the incoherent and interference contributions, and assuming equal lengths $L_1 = L_2 = L$, the Fourier transform of the incoherent part can be written as
\begin{align}
     \mathcal{F}_L \left[P_{\gamma\to X\to \gamma}\right](k) =\;& \sqrt{2\pi}\omega^4\Bigg\{6\sum\limits_n \frac{\beta_n^4}{m_n^8}\left[\frac{\delta (k+4 x_n)}{6}-\frac{2 }{3}\delta (k+2 x_n)+\delta (k)\right]\nonumber\\
     &\nonumber +\sum\limits_{l<n} \frac{\beta_n^2\beta_l^2}{m_n^4 m_l^4}\left\{e^{2iL_w(x_l-x_n)}\left[\delta(k)-2\delta(k-2x_n)+\delta(k-4x_n)-2\delta(k+2x_l) \right.\right.\nonumber\\
     &\left.\left. + \delta(k+4x_l)+\delta(k-4x_n+4x_l) +4\delta(k-2x_n+2x_l)-2\delta(k-4x_n+2x_l)\right.\right.\nonumber\\
     &\left.\left. - 2\delta(k-2x_n+4x_l)\right]\right\}+(x_{n,l} \to -x_{n,l})\Bigg\} \,,
\end{align}
where the $(x_{n,l}\to -x_{n,l})$ indicates adding terms where the replacement $x_n\to -x_n$ is performed to symmetrize with respect to $x_{n,l}$, and
\begin{align}
    &\label{eq:peak-frequency}x_n\equiv \frac{m_n^2}{4\omega}\simeq 1.3\times 10^6 ~\text{m}^{-1}    \left(\frac{\text{eV}}{\omega}\right)\left(\frac{m_n}{\text{eV}}\right)^2 , &\frac{x_n}{2\pi}\simeq 2.0\times 10^5 ~\text{m}^{-1} \left(\frac{\text{eV}}{\omega}\right)\left(\frac{m_n}{\text{eV}}\right)^2 .
\end{align}
For each axion mode, the incoherent contribution therefore produces peaks at $k=\pm 2x_n$ and $k=\pm 4x_n$. This feature is also present in the single axion case.
The interference contributions result in additional peaks at $k=\pm 2(x_n-x_l)$, $k=\pm 4(x_n-x_l)$, $k=\pm (4x_n-2x_l)$ and $k=\pm (2x_n-4x_l)$.

\begin{figure}
    \centering
    \includegraphics[width=0.8\linewidth]{ 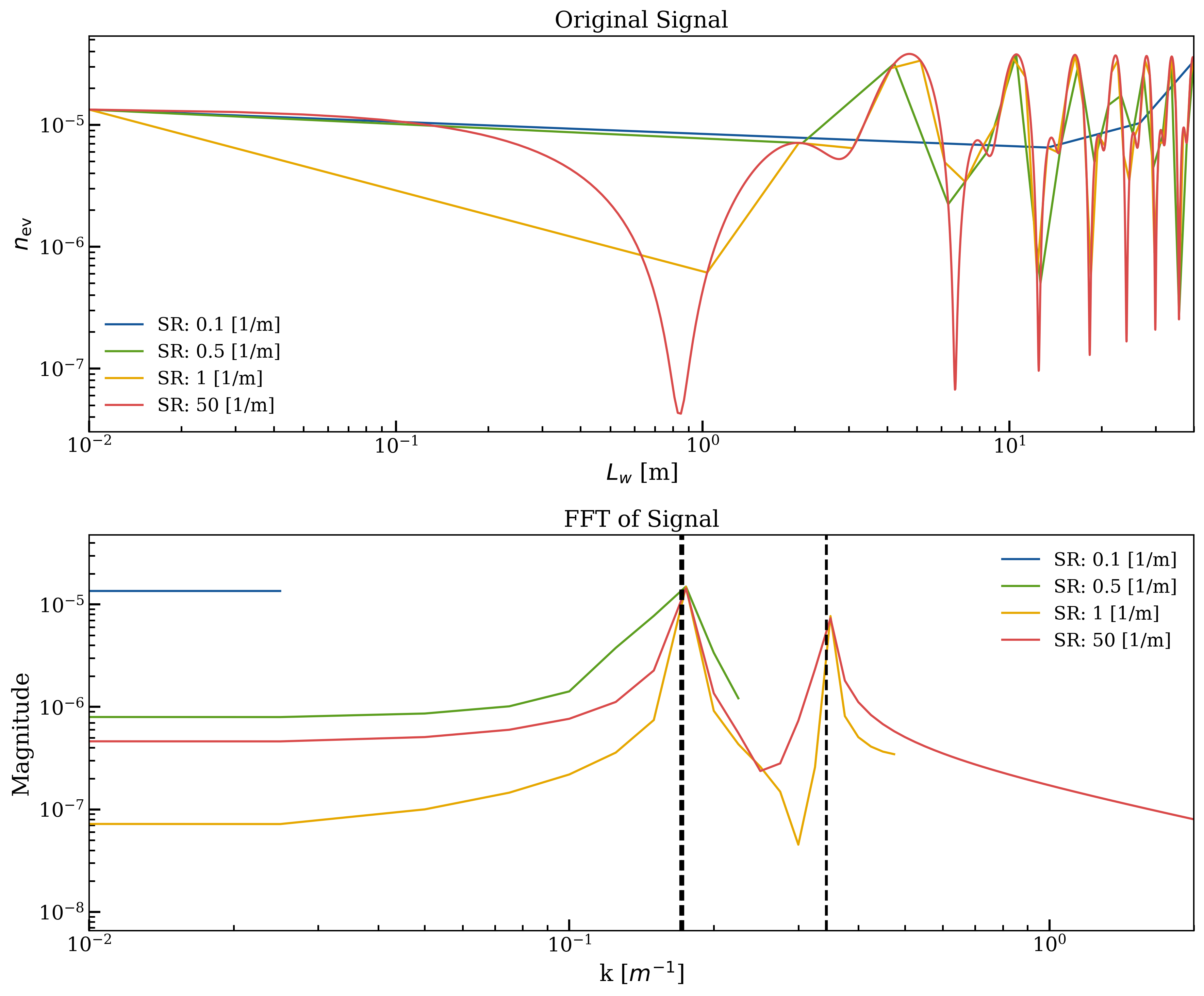}
    \caption{Photon signal obtained using eqs.~\eqref{eq:probability-formula} and~\eqref{eq:signal} for a measurement time of 40 hours per wall-length point scanned and its FFT for different sampling rates (SR) of the wall length considering the KK Maxion benchmark of table~\ref{table-parameters}, scanning wall thicknesses up to $L_w = 40 \,\mathrm{m}$ (all other parameters as in the standard ALPS II setting). For clarity, only the three heaviest (out of a maximum of 100) axion modes are shown. The expected positions of the Dirac delta peaks are indicated by dashed black lines in the lower panel. Owing to the specific structure of the mass spectrum, the feature at $k \simeq 0.2~\mathrm{m}^{-1}$ consists of two overlapping peaks, represented by a thicker dashed~line.\label{fig:FFT_SR-comparison-wall}}
\end{figure}

A complementary approach is to consider the variation of the wall's thickness. Fourier transforming with respect to $L_w$, and defining $x_{ij}\equiv x_i-x_j$, one finds the interference contributions to be
\begin{align}
     \mathcal{F}_{L_w}\left[\cos(2(x_i-x_j)(L+L_w))\right](k)=\sqrt{\frac{\pi }{2}}\left[ e^{2 i L x_{ij}} \delta (k-2x_{ij})+ e^{-2 i L x_{ij}} \delta (k+2 x_{ij})\right].
\end{align}
This yields peaks at $k = \pm 2x_{ij}$. For $N$ axion fields, this procedure gives rise to $N(N-1)/2$ distinct peaks, providing a probe of axion mass splittings and thus a diagnostic of axion multiplicity.
The peak heights give information on the coupling strengths.

Information regarding multiple axions can be extracted by signal
measurements as a function of $L$, or $L_1, L_2, L_w$ in the case of distinct region lengths.
In practice, measurements are performed at a finite number of discrete values of $L_1, L_2$ or $L_w$. As a result, the idealized Dirac delta peak distributions are replaced by peaks whose widths and resolution depend on sampling rates (SR) and on maximum scanned lengths. The SR in particular determines the achievable resolution in axion mass spectroscopy.

In figure~\ref{fig:FFT_SR-comparison-wall} we illustrate the original LSW signal and its fast Fourier transform (FFT) for scans of $L_w$ up to $40 ~\mathrm{m}$ in the KK Maxion benchmark, considering only the three heaviest modes for clarity. The different coloured lines correspond to different SRs of the wall, varying the wall length. Increasing the SR improves the reconstruction of the signal, especially high-frequency contributions, and leads to sharper peaks that better approximate the expected Dirac delta distributions.  The peak structure matches the expectations derived from eq.~\eqref{eq:peak-frequency}, represented as black dashed lines.  For heavier axions, larger SRs are necessary in order to resolve the fast oscillations of the signal. In the example shown in figure~\ref{fig:FFT_SR-comparison-wall}, two of the expected peaks overlap due to the specific structure of the considered benchmark spectrum. Such degeneracies can be lifted experimentally by considering variations in the lengths of the active magnetic field regions, as its Fourier transform is sensitive to the absolute axion mass scale and not just mass differences. Generically, larger maximum values of the scanned experimental lengths are needed to resolve smaller masses and mass differences.

\section{Helioscopes and solar axion detection}
\label{sec:helioscopes}

Helioscopes can probe axions produced in the Sun through their conversion into photons in a laboratory magnetic field~\cite{Sikivie:1983ip}.\footnote{Axions may also be converted back into X-rays in the Sun's own magnetic field, allowing for a detection in suitable telescopes~\cite{Ruz:2024gkl}. But this does not affect the signal of interest in the present~section.} While the axion source is not experimentally controlled, the solar axion flux can be significant and is theoretically well understood, enabling stringent constraints on the axion-photon coupling from both direct searches and stellar energy-loss arguments~\cite{Raffelt:1985nk,Vinyoles:2015aba}. In the context of multiple axions, helioscope limits have previously been studied
in anarchical models~\cite{Chadha-Day:2021uyt,Chadha-Day:2023wub}, including reinterpretations of CAST data~\cite{CAST:2013bqn}.
There, the aim was to set limits from the overall flux of regenerated photons. Here, we go beyond this, aiming to also differentiate between a signal originating from a single axion and that of multiple axions. To do this, we include coherence and interference effects both in the Sun as well as in the laboratory part of the experiment. We also consider the spectral flux and investigate features arising from multiple axions. We also explore the use of a buffer gas to gain additional clear information on a multiple axions origin.
For theoretical models we focus on the structured benchmark models introduced in section~\ref{sec:models}.
In particular, we examine the impact of a nearly degenerate axion mass spectrum, as can arise, e.g., in axiverse stringy axion scenarios, and discuss how helioscope measurements could provide diagnostic information distinguishing single and multiple axion frameworks.

\enlargethispage*{\baselineskip}\relax
The regeneration stage of helioscope experiments, such as CAST and IAXO~\cite{CAST:2004gzq,CAST:2013bqn,CAST:2017uph,Armengaud:2014gea,IAXO:2019mpb,IAXO:2020wwp}, is analogous to that of LSW experiments and thus allows direct application of the results derived in section~\ref{sec:LSW-pheno}. A key difference from LSW experiments lies in the higher axion energies, which are typically in the keV range for solar axions. Combined with the shorter length scales of near-future helioscopes, this grants sensitivity to larger axion masses than those expected to be accessible in LSW setups. Consequently, signatures of multiple axions appear at correspondingly larger mass scales. In this section, we therefore adopt the helioscope benchmark parameters listed in table~\ref{table-parameters}, obtained by rescaling the LSW benchmark masses $m_0$, $m_a$, and $\mu_1$ by a factor of $10^3$. For the KK Maxion benchmarks, we further increase the decay constants by a factor of $\sqrt{10}$ relative to the KK ALP case in order to obtain signals of comparable magnitude.

As a reference, for helioscope setup we consider experimental parameters akin to the proposed IAXO experiment~\cite{IAXO:2019mpb}.

\subsection{Differential photon flux calculation}
Axions can be produced in the Sun through several mechanisms,
including the Primakoff process, Compton scattering and Bremsstrahlung (cf., e.g.~\cite{Redondo_2013}).
In the present analysis, we focus exclusively on axion-photon couplings, such that Primakoff production provides the dominant contribution. For the purposes of this qualitative study, we employ an analytical approximation for the Primakoff spectrum~\cite{Irastorza_2011}
\begin{align}\label{hel::fluxes}
    \left(\frac{d\phi}{d\omega}_\text{Prim}\right)_{n}=\eta^{2}_{n}=2.0\times10^{18}\left(\frac{c_n g_{a\gamma}}{10^{-12}~\text{GeV}^{-1}}\right)^2\left(\frac{\omega}{\rm{keV}}\right)^{2.450}e^{-0.829\omega/{\rm{ keV}}}\text{yr}^{-1}\text{m}^{-2}\text{keV}^{-1}  .
\end{align}
A more detailed quantitative analysis could be carried out using improved numerical calculations along the lines of~ref.~\cite{Hoof:2021mld}. Such an analysis should also include possible contributions from other production mechanisms in the Sun, notably plasmon conversion in the internal magnetic fields of the Sun~\cite{Caputo:2020quz,OHare:2020wum,Guarini:2020hps}. We do not include these contributions in our present analysis. While this will affect the quantitative results, our main strategy of searching for features under variation of experimental parameters and in the spectrum remains unaffected.

The photon flux in an axion helioscope can now be computed by combining the production spectrum with the axion-photon conversion probability for regeneration within the detector magnet. This includes accounting for the propagation from the solar production region to the helioscope over a distance $L_{\text{Sun}}$, and using the results of section~\ref{sec:LSW-pheno}, treating $L_{\rm Sun}$ effectively as a very extended wall $L_w$.
The resulting differential photon flux per unit energy and   area, assuming the relativistic limit for simplicity (see appendix~\ref{app:eikonal} for the full expression), can be written as
\begin{align}\label{hel::prob}
\frac{d^{2}F}{d\omega dA}=\;&\sum\limits_{n=1}^N \eta_n^2 \frac{4\beta_n^2\omega^2 }{m_n^4}\sin^{2}\left(\frac{L_\text{mag}m_n^2}{4\omega}\right
)\\
&+2\sum\limits_{m<n}^N\eta_n\eta_m\frac{4\beta_n\beta_m\omega^2}{m_n^2m_m^2}\sin\left(\frac{L_\text{mag}m_n^2}{4\omega}\right)
\sin\left(\frac{L_\text{mag}m_m^2}{4\omega}\right)\cos\left(\left(L_\text{Sun}+\frac{L_\text{mag}}{2}\right)\frac{m_n^2-m_m^2}{2\omega}\right) \nonumber ,
\end{align}
where $\beta$ is the product of the magnetic field strength on the regeneration region and the axion-photon coupling. Here, $L_\text{mag}$ is the length of the magnet in the laboratory part of the experiment. The $\eta_n$ effectively represent\footnote{Essentially, the quantities $\eta_n$ are proportional to the moduli of the axion production~amplitudes.} the square roots of the axion production rates of eq.~\eqref{hel::fluxes}. Here, we neglect possible reconversion effects inside the Sun, as they are subdominant.

Similar to the LSW case, the second line of eq.~\eqref{hel::prob} encodes coherence and decoherence effects arising from interference between different axions. Depending on the mass splittings, this contribution can either enhance or suppress the total signal. In ref.~\cite{Chadha-Day:2023wub}, this interference term was averaged out by integrating over the axion energy spectrum. This is justified when the mass splittings satisfy $\Delta m^2 \gtrsim 10^{-12} \mathrm{eV}^2$. In that regime, the rapidly oscillating cosine averages to zero. However, this approximation does not always apply to our scenarios. In particular, the stringy axion models feature a number of closely spaced axion masses. Moreover, we are interested in energy-resolved detection, which necessitates a more careful treatment of the interference term.

We note that eq.~\eqref{hel::prob}, implicitly assumes that all axions are produced at a fixed distance $L_{\text{Sun}}$ from the detector, which is a simplification. In reality, solar axion production occurs over an extended region, with approximately $\sim90\%$ of the flux originating from within the inner $\sim30\%$ of the solar radius, corresponding to $R_{90\%} \sim 2 \times 10^{5}~\mathrm{km}$~\cite{Hoof:2021mld}. For sufficiently large mass splittings, integrating over this production region leads to a rapid dephasing of the cosine term, effectively suppressing the interference contribution. In this case, only the incoherent sum in the first line of eq.~\eqref{hel::prob} remains relevant.

For small mass splittings, however, the phases associated with axions produced at different locations in the Sun can remain partially coherent. A fully consistent treatment would then require an explicit integration over a solar model in order to properly account for the contributions of the cosine term. For the logarithmically distributed mass spectra characteristic of axiverse stringy scenarios\footnote{In the other benchmark models considered, mass splittings are sufficiently large that this issue does not~arise.} only a small fraction of axion pairs fall into this regime. Most axion mass pairs either have phases much smaller than unity, effectively setting the cosine to one, or much larger than unity, averaging it to zero. Hence, in both cases, the precise production distance becomes irrelevant.

\begin{figure}
    \centering  \includegraphics[width=0.55\linewidth]{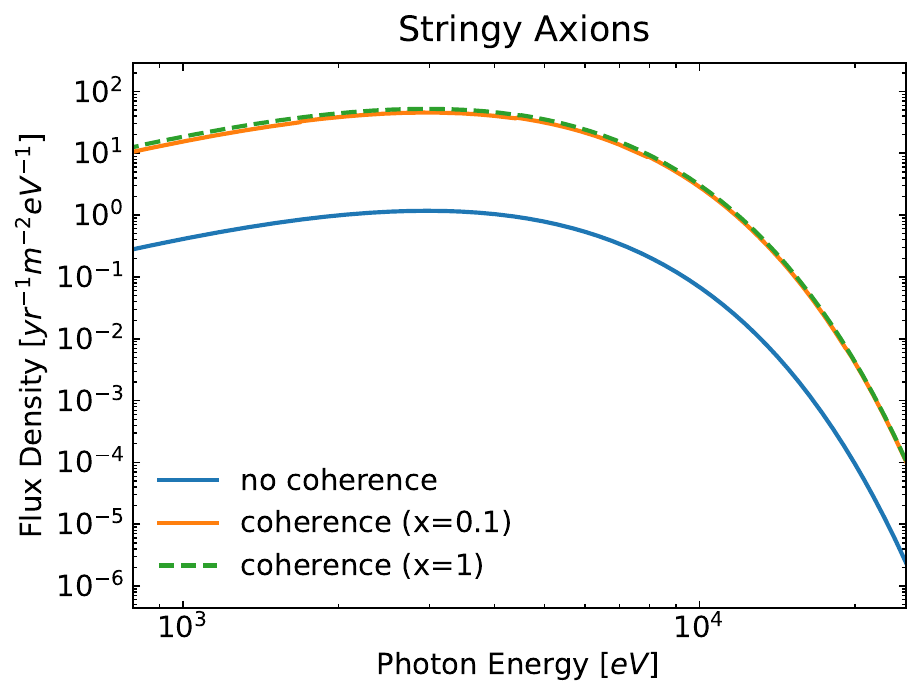}
    \caption{Differential photon flux at a terrestrial detector for the stringy axion benchmark model of table~\ref{table-parameters}, comparing different treatments of coherence effects. Neglect of coherence entirely (blue), inclusion of coherence for axion pairs with $\Delta m^2 < 0.1 \times 2\pi(\omega/L_{\rm solar})$ (orange), and inclusion of coherence for axion pairs with $\Delta m^2 < 2\pi  (\omega/L_{\rm solar})$ (green) are considered. The results are shown for a regeneration length $L_\text{mag} = 20~\mathrm{m}$ and a magnetic field strength $B = 9~\mathrm{T}$. We have taken $L_{\rm solar} =R_{90\%}= 2 \times 10^{8}~\mathrm{m}$ as a representative value.\label{fig:comp_coherence}}
\end{figure}

We therefore adopt a simplified prescription in which the interference term is retained only for axion mass pairs satisfying
\begin{equation}
    \frac{(m_n^2-m_m^2) L_{\rm solar}}{2\pi \omega}\lesssim x\simeq 1 ,
\end{equation}
where $L_{\rm solar}$ indicates the rough size of the production region $L_{\rm solar} \sim R_{90\%}$.
Above this threshold, we set the coherent term to zero, and below it we set it to unity.
In the regime where coherence is preserved, the photon flux effectively becomes proportional to the square of the sum of the individual conversion amplitudes, rather than the sum of their squares. This leads to an overall enhancement of emission by a factor of $N$ relative to the incoherent case, analogous to the coherent LSW regime.\footnote{For the emission, we could adopt the same argument as in eqs.~\eqref{eq:cohere1} and~\eqref{eq:cohere2}.}

In figure~\ref{fig:comp_coherence}, we show that varying numerically the point $x$ at which we make the coherence cut by an order of magnitude does not affect the resulting differential photon flux considerably. However, neglecting coherence altogether leads to sizeable deviations.
In the following, we use $x=1$.

\subsection{Total photon flux as a function of experimental length}

\begin{figure}
    \centering
    \begin{subfigure}{0.495\textwidth}
        \includegraphics[width=\linewidth]{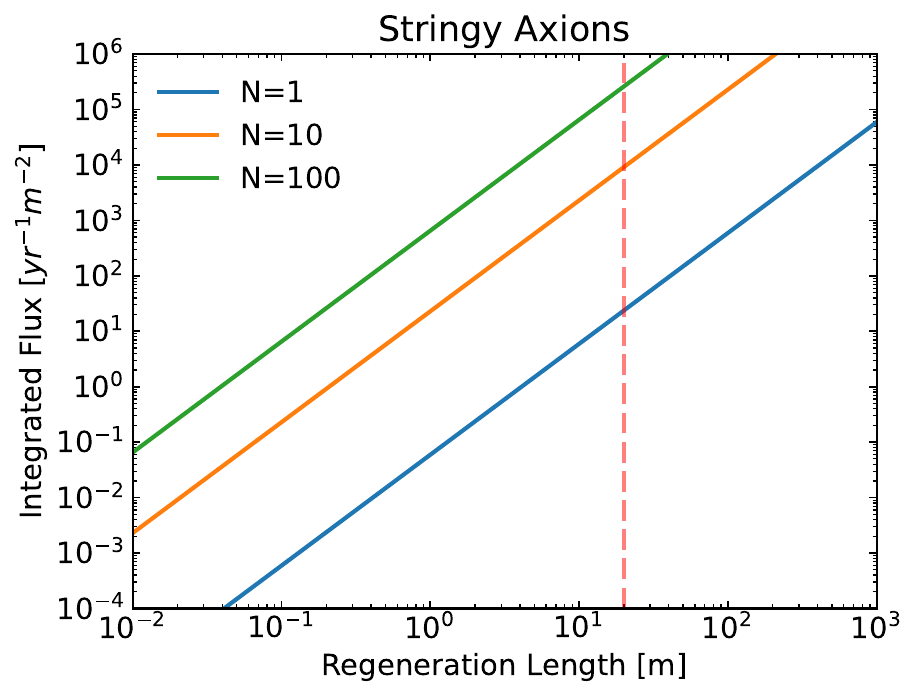}
    \end{subfigure}\hfill
    \begin{subfigure}{0.495\textwidth}
        \includegraphics[width=\linewidth]{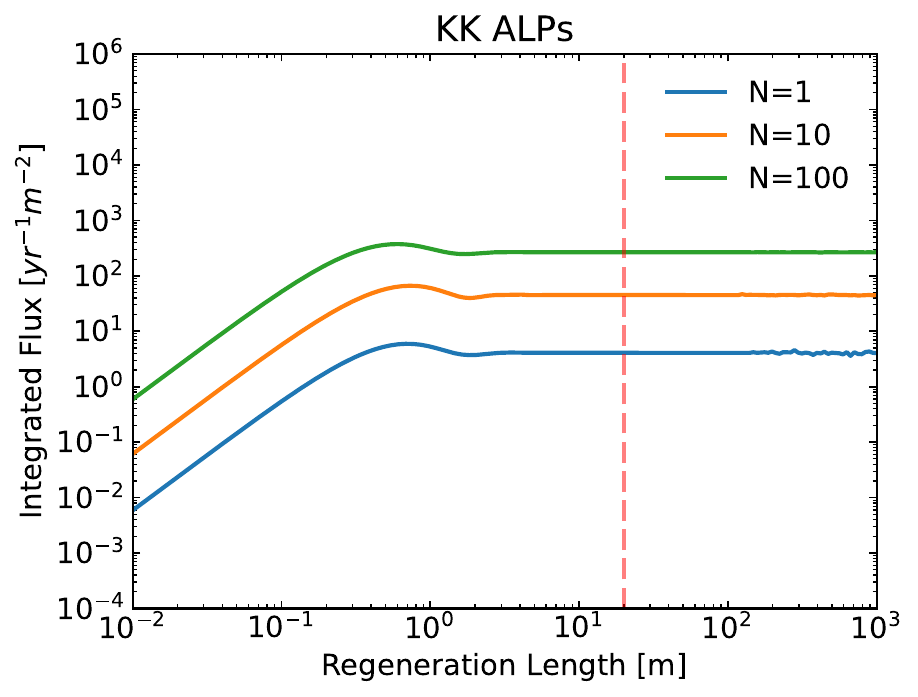}
    \end{subfigure}
\\[8pt]
    \begin{subfigure}{0.495\textwidth}
        \includegraphics[width=\linewidth]{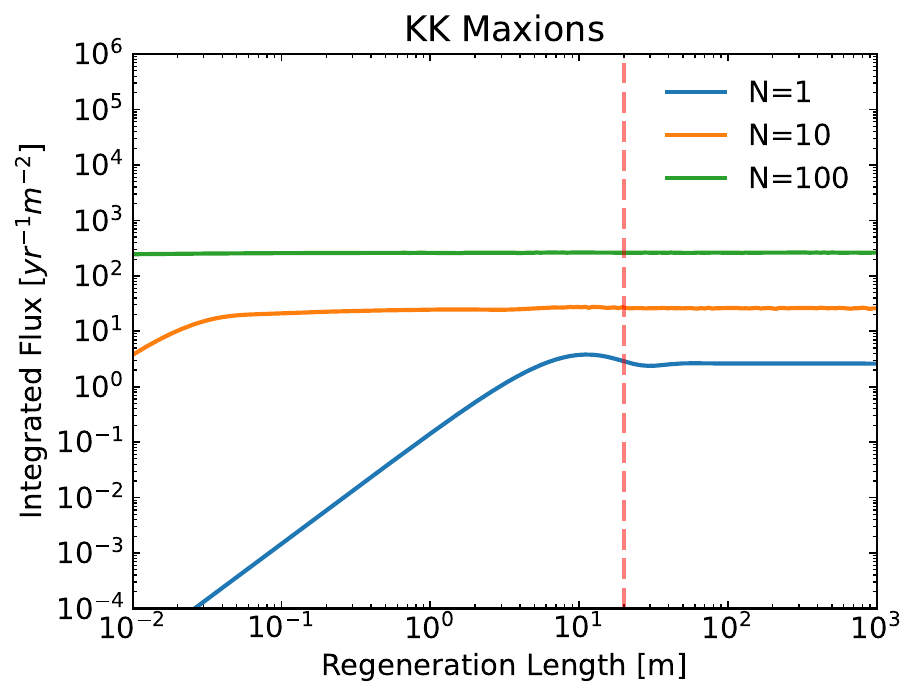}
    \end{subfigure}
    \caption{Energy integrated axion-induced photon flux in helioscope experiments as a function of the regeneration length. The magnetic field strength is fixed to $B = 9~\mathrm{T}$. {The red dashed line marks the regeneration length expected in IAXO~\cite{Armengaud:2014gea}}. Benchmark model parameters are specified in table~\ref{table-parameters}.\label{fig:heliosignals_models_length}}
\end{figure}

We now consider the energy $(\omega)$ integrated flux obtained from eq.~\eqref{hel::prob}. As in the LSW case, it is instructive to examine the dependence of the signal on the length of the regeneration region. The resulting behaviour is shown in figure~\ref{fig:heliosignals_models_length} for regeneration lengths between $0.01~\text{m}$ and $1000~\text{m}$, and the benchmark models defined in table~\ref{table-parameters}, considering $N = 1, 10, 100$. The shown range of regeneration lengths has been chosen to exceed the foreseen specifications of IAXO ($\sim20~\text{m}$~\cite{Armengaud:2014gea}) by up to approximately one order of magnitude at the largest values, and is included here for reference.

We begin with the stringy axion benchmark, shown in the upper left panel of figure~\ref{fig:heliosignals_models_length}. Similar to what we observed before, for short regeneration lengths and nearly mass-degenerate light axions, the signal exhibits the expected $\sim N^2$ scaling characteristic of coherent conversion, as illustrated by the comparison of the blue and orange curves. As we include successively heavier modes, the system transitions into the incoherent regime, and the scaling reduces to $\sim N$. For our chosen range of $L$, this happens between $N=10$ and $N=100$. Since the regeneration length is much shorter than the solar production region, the signal is dominated by the lightest axion modes, and no pronounced oscillatory features appear over the range of lengths considered.

The situation differs for the KK ALPs and KK Maxions benchmarks shown in the right and lower panels. In these cases, oscillations within the regeneration region become relevant as the length increases. However, because the oscillation length depends on the photon energy, integrating over the solar axion spectrum effectively averages over these oscillations, rendering the corresponding features less distinct. For sufficiently long regeneration lengths, the larger characteristic axion masses lead to a saturation of the signal.

While measurements at multiple regeneration lengths can, in principle, still provide information on axion multiplicity, the discriminatory power of energy-integrated observables is limited. For this reason, we next turn to energy-resolved fluxes and discuss the use of a buffer gas as a means to enhance sensitivity to spectral features of multiple axions.

\subsection{Spectral photon flux}

As noted above, the broad energy spectrum makes the features associated with multiple axions less pronounced when regeneration length variations are considered.  However, this broad spectrum can be exploited by noting that the trigonometric factors in eq.~\eqref{hel::prob} depend explicitly on the photon energy $\omega$. Already in the single axion scenario, it was observed that this energy dependence can be used to infer the axion mass from spectral features~\cite{CAST:2008ixs,Jaeckel:2018mbn,Dafni:2018tvj}.

\begin{figure}
    \centering
    \begin{subfigure}{0.495\textwidth}
        \includegraphics[width=\linewidth]{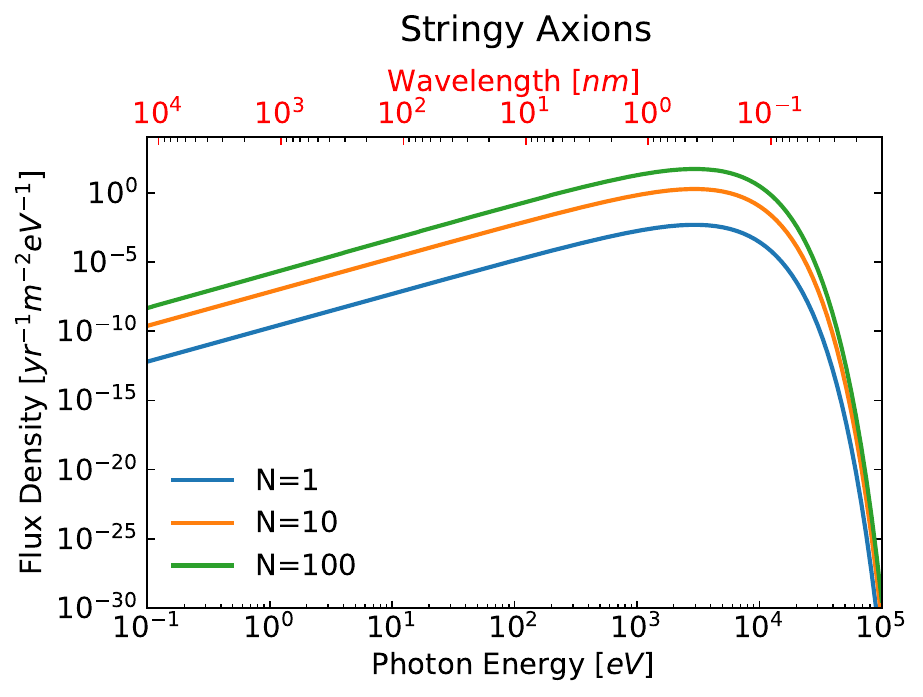}
    \end{subfigure}\hfill
    \begin{subfigure}{0.495\textwidth}
        \includegraphics[width=\linewidth]{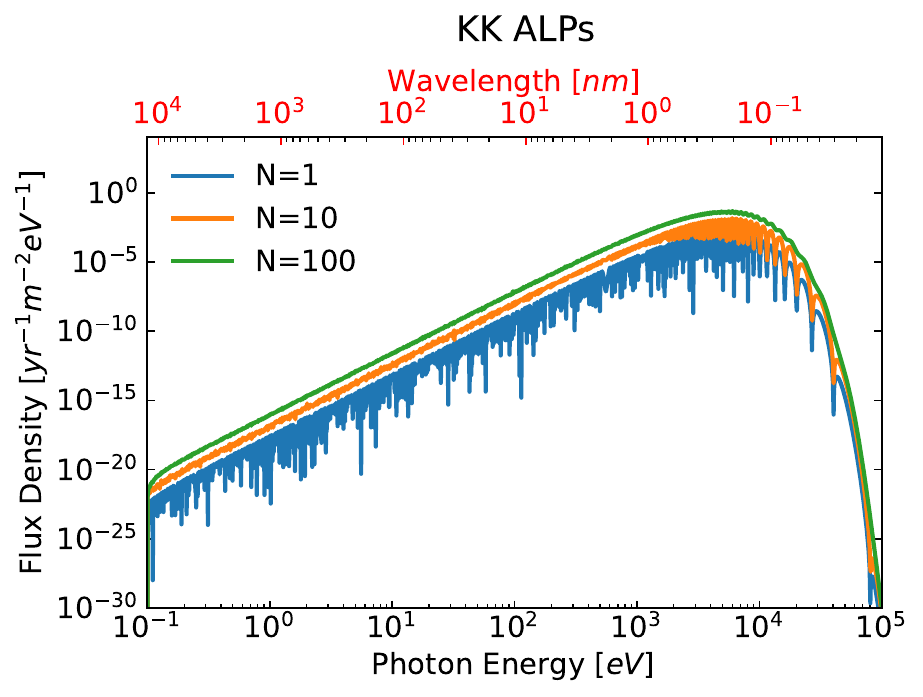}
    \end{subfigure}
    \\[8pt]
    \begin{subfigure}{0.495\textwidth}
        \includegraphics[width=\linewidth]{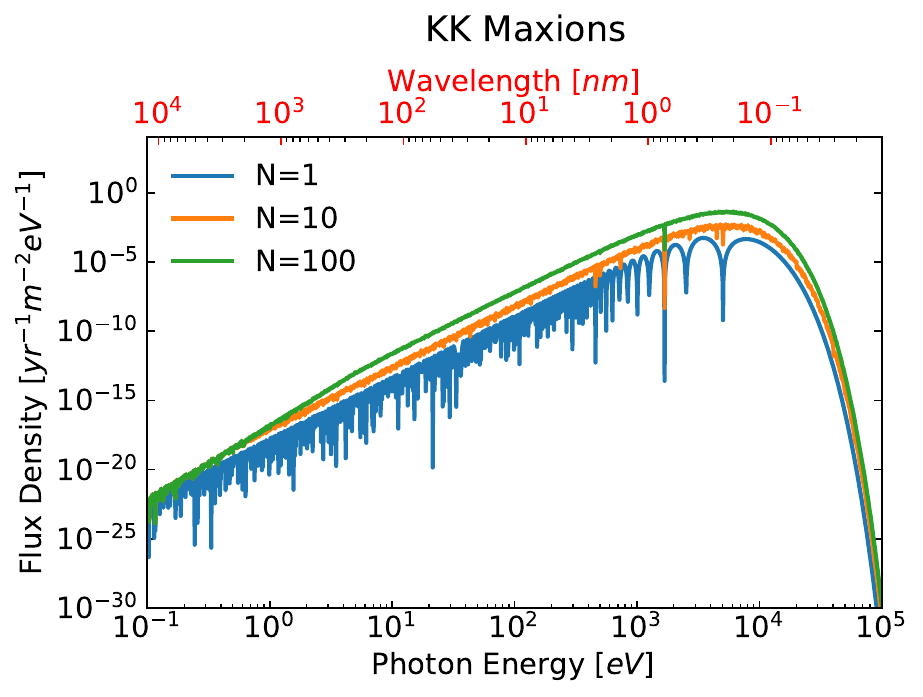}
    \end{subfigure}
    \caption{Energy-resolved regenerated photon flux in helioscope experiments as a function of the photon energy $\omega$. The regeneration length is taken to be $L = 20\, \mathrm{m}$ and the magnetic field strength to $B = 9\, \mathrm{T}$. Axion benchmark parameters are given in table~\ref{table-parameters}.\label{fig:heliosignals_models_frequency}}
\end{figure}

In figure~\ref{fig:heliosignals_models_frequency}, we show the resulting energy spectra\footnote{We note that at low frequencies the spectrum might be quantitatively impacted by including plasmon production~\cite{Caputo:2020quz,OHare:2020wum,Guarini:2020hps}. This is straightforward to include in a more detailed~analysis.} of reconverted photons in a helioscope.\footnote{S.M. would like to thank M.~Giannotti for helpful~comments.} In the single axion scenario (blue curves), distinct spectral modulations appear once the axion mass is sufficiently large for oscillations to develop within the regeneration region. This behaviour is visible for the KK ALPs and KK Maxions benchmarks, whereas in the stringy axion scenario, the lightest axion is chosen sufficiently light that oscillations do not occur over the relevant length scale. When additional axions are included, these spectral features are substantially reduced. The superposition of contributions from multiple mass eigenstates leads to an effective averaging of oscillatory structures. An apparent exception arises in the KK Maxion benchmark, where a near half-integer relation among masses (see table~\ref{table-parameters}) produces aligned oscillation patterns. In practice, however, the heavier modes induce significantly more rapid oscillations, resulting in ``narrower'' spectral zeros. These features are not resolved at the numerical resolution employed here, which already exceeds realistic experimental energy resolutions.

Despite this averaging, the low-energy behaviour of the spectrum remains sensitive to multiple axions. In particular, when all contributing axions lie in the oscillatory regime, the conversion probability scales as $\omega^2$ according to eq.~\eqref{hel::prob}, leading to a noticeably steeper low-energy slope relative to scenarios dominated by non-oscillating modes. This feature persists even in the presence of multiple axions and may provide a complementary handle on multiple axions in energy-resolved measurements.

\subsection{In-medium photon effects and buffer gas}

A further experimentally tunable parameter that can be used to probe multiple axion scenarios is the pressure and temperature of a buffer gas introduced into the regeneration region. In single axion searches, varying the gas density has long been exploited to extract information about the axion mass by tuning the photon dispersion relation~\cite{CAST:2008ixs,Jaeckel:2018mbn}.

\begin{figure}
    \centering
    \begin{subfigure}{0.495\textwidth}
        \includegraphics[width=\linewidth]{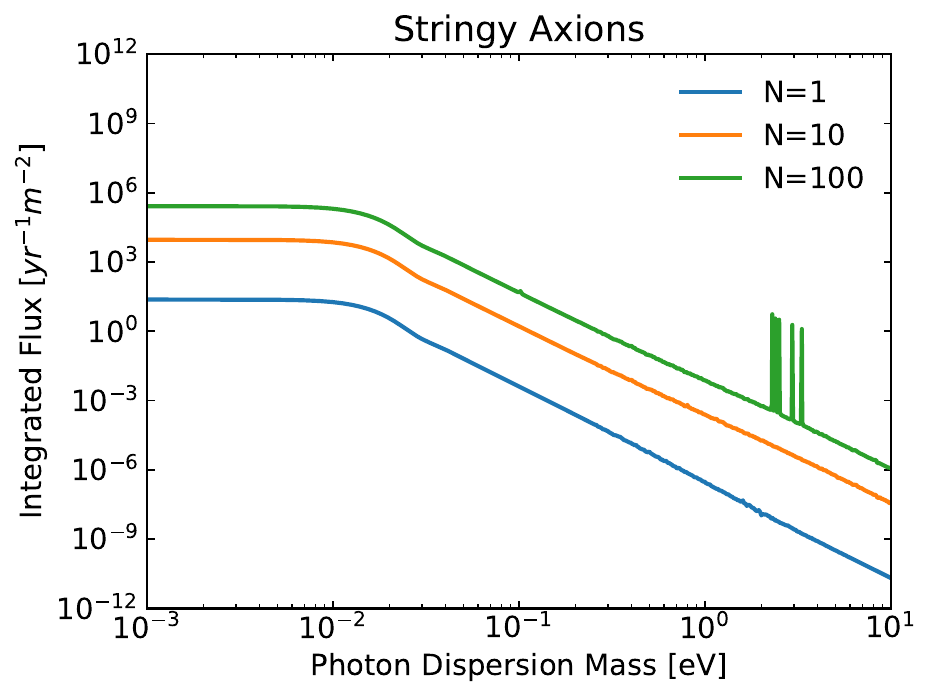}
    \end{subfigure}\hfill
    \begin{subfigure}{0.495\textwidth}
        \includegraphics[width=\linewidth]{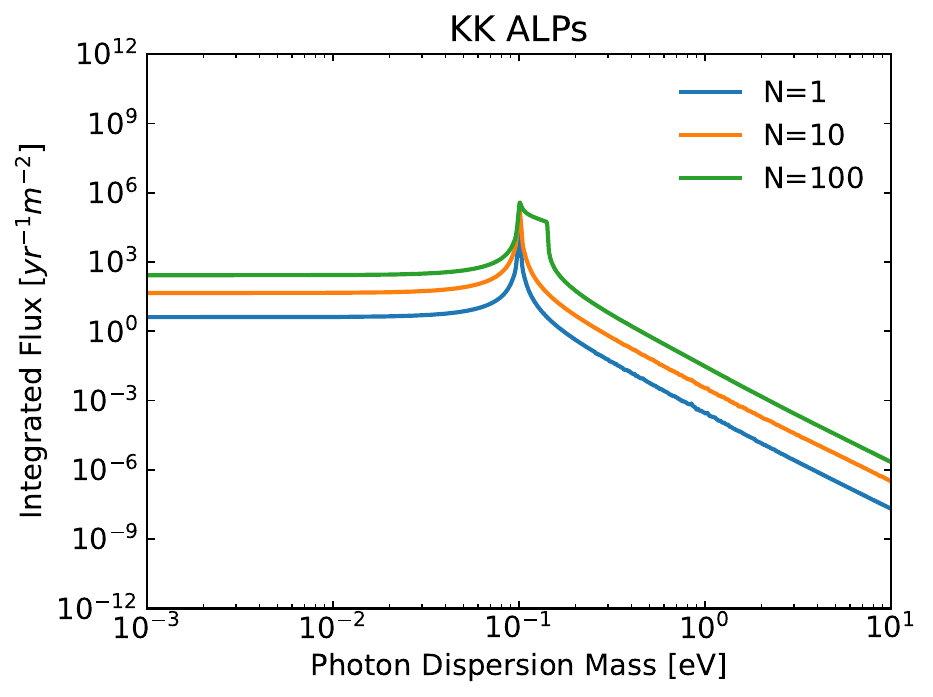}
    \end{subfigure}
    \\[8pt]
    \begin{subfigure}{0.495\textwidth}
        \includegraphics[width=\linewidth]{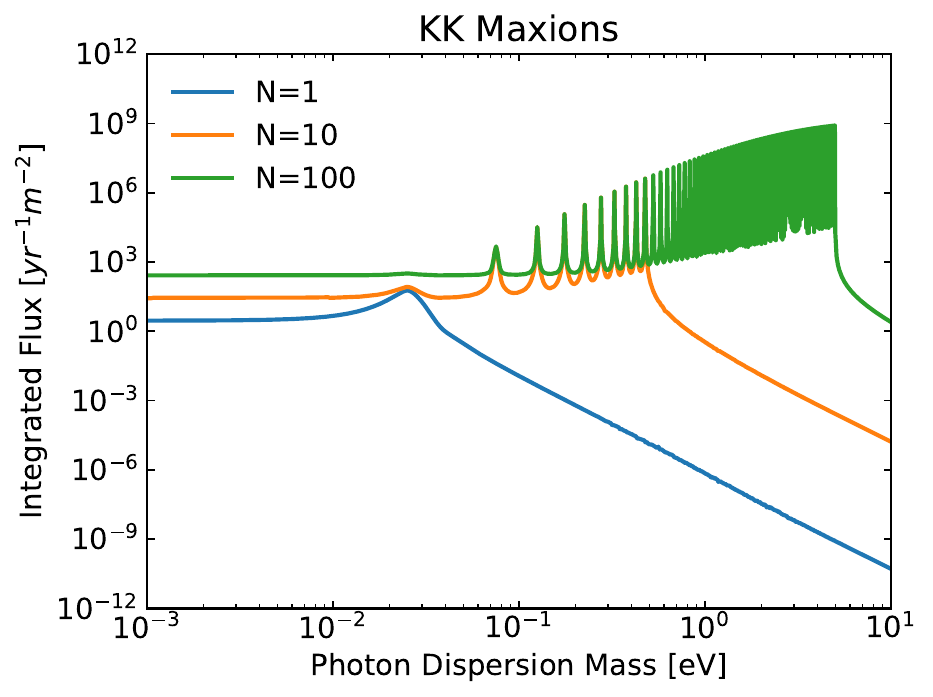}
    \end{subfigure}
    \caption{Energy-integrated photon flux in a helioscope experiment with $B = 3~\mathrm{T}$ and $L = 100~\mathrm{m}$ as a function of the buffer gas pressure in the regeneration region at room temperature (with $T = 293.15~\mathrm{K}$). See helioscopes lines in table~\ref{table-parameters} for parameter~details.\label{fig:heliosignals_models_resonant}}
\end{figure}

The presence of a buffer gas effectively renders the in-medium photons massive, but also leads to photon absorption. Assuming a $^4\text{He}$ buffer gas, we adopt the parametrization (cf., e.g.~\cite{maril8effective,Jaeckel:2018mbn}),
\begin{align}\label{eq:pressure}
    &m_\gamma \simeq \sqrt{0.02\frac{(p/\text{mbar})}{(T/{\text{K}})}} ~\text{eV} ,
    &\Gamma\simeq 0.29\frac{(p/\text{mbar})}{{(\omega/\text{keV})}^{3.1} (T/{\text{K}})} ~\text{m}^{-1} .
\end{align}
Here, $p$ and $T$ denote the gas pressure and temperature, respectively, and $\Gamma$ is the photon absorption coefficient as before. The helioscope conversion probability in eq.~\eqref{hel::prob} must then be modified analogously to the in-medium expression given in eq.~\eqref{eq:P_medium}. Below, we assume a constant temperature of $T = 293.15~\mathrm{K}$ for reference.

In figure~\ref{fig:heliosignals_models_resonant} we display the resulting energy-integrated photon flux scanned over the gas pressure variation, equivalently the effective photon masses, for each of the benchmark models introduced in section~\ref{sec:models}.
The qualitative behaviour of the photon flux is consistent with expectations. Whenever the effective photon mass matches an axion mass eigenvalue, the conversion probability is resonantly enhanced. The distinct spectral structures of the benchmark models are clearly visible in figure~\ref{fig:heliosignals_models_resonant}. In the stringy axion case, the widely spaced masses give rise to well-separated resonance peaks. By contrast, the nearly degenerate KK ALPs spectrum produces a series of closely spaced resonances around $m_\gamma \simeq 10^{-1}~\mathrm{eV}$, which collectively appear as a single broad feature. The KK Maxions benchmark exhibits an intermediate behaviour, with partially overlapping but still distinguishable resonances. Varying the gas pressure and temperature, therefore, provides a powerful experimental handle to probe the axion mass spectrum in multiple axion scenarios.

While we focus here on solar axions as a continuous relativistic source for helioscopes, other classes of relativistic axion fluxes can be qualitatively distinct. Transient astrophysical events, such as axion star bosenovae, compact-object mergers, or supernovae can produce short-duration bursts of relativistic axions that can be probed by laboratory and satellite experiments~\cite{Grifols:1996id,Brockway:1996yr,Payez:2014xsa,Eby:2021ece,Arakawa:2023gyq,deGiorgi:2024pjb,Arakawa:2025hcn}.
More generally, the superposition of many historic transient events can give rise to an effectively diffuse axion background~\cite{Eby:2024mhd}.
A notable source of astrophysical signals of relativistic axion emission are core-collapse supernovae and subsequent conversion in cosmic magnetic fields. This has also been discussed, both for transients and diffuse backgrounds~\cite{Grifols:1996id,Brockway:1996yr,Payez:2014xsa,Raffelt:2011ft,Calore:2020tjw}.
For transient bursts, the superposition of axion waves can retain coherence for nearby sources and very similar axion masses, whereas for a diffuse background generated by many and distant sources, we expect phase coherence to be lost. A more quantitative analysis for the multiple ALP case will require taking into account the source spectra as well as coherence and interference effects between the different mass states. Nevertheless, the axion-photon conversion formalism we derived also applies, with observable signatures primarily determined by the temporal and spectral coherence of the incoming axion flux.

\section{Haloscopes and axion dark matter detection}
\label{sec:haloscopes}

Haloscopes provide excellent mass resolution for the detection of DM axions through resonant conversion in a background magnetic field. In multiple axion scenarios with sufficiently separated axion masses, this would manifest as multiple distinct resonance lines, one associated with each species. However, a detection at a single resonance frequency would not by itself determine whether the signal originates from one dominant axion or from multiple axions and with others yet undetected, nor does it uniquely predict where additional axions might appear in other parts of frequency space. In the following, we comment on general qualitative features, considering haloscopes searches such as ADMX~\cite{ADMX:1998pbl,ADMX:2025vom}, CAPP~\cite{Lee:2019mfy,CAPP:2020utb,CAPP:2024dtx}, HAYSTAC~\cite{HAYSTAC:2018rwy,HAYSTAC:2024jch}, ORGAN~\cite{McAllister:2017lkb,Quiskamp:2024oet}, QUAX~\cite{Alesini:2020vny,QUAX:2024fut}, or proposed experiments such as MADMAX~\cite{Caldwell:2016dcw,MADMAX:2019pub} and DMRadio-$m^3$~\cite{DMRadio:2022pkf} (see also~\cite{Ringwald:2024uds} for a brief review).

Depending on the model and cosmological history of the Universe, one or more axion species could constitute a sizable fraction of DM and contribute to its local density. Notably, those could be produced~\cite{Arvanitaki:2009fg,Acharya:2010zx,Luu:2018afg,Reig:2021ipa,Murai:2023xjn,Murai:2024nsp,Li:2025uwq} through the misalignment mechanism~\cite{Preskill:1982cy,Abbott:1982af,Dine:1982ah,Arvanitaki:2009fg,Arias:2012az} or through the decay of topological defects\footnote{See~refs.~\cite{Davis:1985pt,Davis:1986xc,Vilenkin:1986ku,Harari:1987ht} for some of the earlier studies of single axion production through the decay of topological~defects.}~\cite{Benabou:2023npn,Lee:2024toz}.
In the non-relativistic regime, each contributing axion field can be approximately treated as a coherently oscillating background
\begin{equation}
    a_n(t)\simeq \frac{\sqrt{2\rho_n}}{m_n}\cos(m_n t +\delta_n) ,
\end{equation}
where $\rho_n$ is the local energy density of each species and $\delta_n$ are random phases.
The values of $\rho_n$ depend on the underlying model and the cosmological production history.

In the presence of a background magnetic field $\Vec{B}_b$, axion field oscillations can induce an effective periodic current
\begin{equation}
    \Vec{J}=-\Vec{B}_b~\sum\limits_{n=1}^N g_{a_n\gamma}\left(\frac{\partial a_n}{\partial t}\right) ,
\end{equation}
whose frequency is dictated by the axion mass, and which can then be detected.
Many haloscope experiments are designed to resonantly detect such currents and are therefore sensitive to axion masses through scans of the detector's resonant frequency, achieved by varying the experimental setup. In multiple axion scenarios this generically implies the presence of multiple resonant responses associated with the different axion species, provided that their mass splittings $(m_{i}-m_{j})/m_{i}$  exceed the intrinsic fractional linewidth associated with the axion signal, set by the Galactic velocity dispersion\footnote{We consider a standard DM halo model with~$v \sim 10^{-3}$.} $\Delta \omega/\omega \sim v^2 \sim 10^{-6}$    and the experimental bandwidth, related to quality factor. For smaller mass splittings, the corresponding resonances overlap and appear as a single broadened or otherwise modified spectral feature rather than as distinct peaks. Coherent enhancement can occur only for very degenerate axions, when $(m_{i}-m_{j})/m_{i}\lesssim 10^{-6}-10^{-3}$ depending on the quality factor of the resonant experimental setup. As an example, such near-degeneracies could be realized in the KK ALP scenario when $\mu_1/m_0 \lesssim 10^{-3}-10^{-1}$.

There are several further advantages to haloscope searches in this context.
First, the axion mass is directly measured with high precision through the resonant frequency of the detector. This contrasts with LSW experiments and helioscopes, where precise mass determination for light axions might be challenging.
Second, since haloscopes do not require additional axion production beyond the cosmological relic abundance the resulting signal typically depends on a lower power of  weak axion-photon couplings. Thus, haloscopes could be more sensitive to smaller couplings.
Third, a haloscope detection, if confirmed by an LSW or helioscope experimental signal, can provide valuable information on the present-day abundance of such axion species.

To briefly illustrate the complementarity, we consider a set of relic energy densities $\{\rho_n\}_n$ for the different axion species. The haloscope signal-to-noise ratio~(SNR) for a given axion mode is then expected to be proportional to
\begin{equation}
    \text{SNR}_n\propto g_{a_n\gamma}^2 \rho_n .
\end{equation}
The exact prefactor depends on the considered experimental setup. This implies that haloscopes are sensitive only to the product of local energy densities and couplings. This is not the case for LSW experiments, where the axion flux is artificially produced and is under experimental control. Consequently, once the axion mass is known, an LSW experiment can, in principle, determine $g_{a_n\gamma}$, regardless of the cosmological abundance. While challenging in practice, a haloscope signal can thus guide the design of a targeted LSW experiment sensitive to axion detection even if the coupling is small~\cite{Hoof:2024gfk}.

Without independent knowledge of the model-dependent coefficients $c_i$, the degeneracy between the energy densities and the couplings cannot be broken. Nevertheless, within a given theoretical framework, measurements of the relative signal strengths that depends on the axion abundance
\begin{equation}
    \frac{\text{SNR}_i}{\text{SNR}_j}=\frac{g_{a_i\gamma}^2 \rho_i}{g_{a_j\gamma}^2 \rho_j}=\frac{c_i^2}{c_j^2}\frac{\rho_i}{ \rho_j}
\end{equation}
can substantially constrain the viable cosmological production scenarios. A number of studies considering multiple axion cosmology along these directions already exist~\cite{Mack:2009hs,Dienes:2011ja,Dienes:2011sa,Dunsky:2025sgz,Dessert:2025yvk,Asadi:2025cvm}. Given the significant model dependence we do not delve into further details here.

\section{Conclusions}\label{sec:conclusions}

While minimal extensions of the Standard Model are often attractive in their simplicity, it is far from certain that Nature has chosen this path. This is highlighted in the case of axions and axion-like particles. Motivated theories, notably including extra-dimensional and string theory-based ones, generically feature a multitude of axion fields. In this work, we have focused on investigating the experimental manifestations of multiple axions in laboratory-based searches, with particular emphasis on highly controllable light-shining-through-a-wall~(LSW) setups, while also illustrating their complementarity with helioscopes probing solar axions and haloscopes targeting relic dark matter~(DM) axion populations.

Our analysis was guided by two central questions. First, does the presence of multiple axions generically enhance experimental sensitivity compared to a single axion with comparable couplings? Second, can experimental observations distinguish between single and multiple axion scenarios and potentially extract additional information, such as mass splittings?

To address these questions, we discuss changes in the signals in LSW experiments, helioscopes and haloscopes under variation of experimental parameters and identifying suitable observables. For concreteness, we consider a number of theory-motivated benchmark scenarios with multiple axions. While specific features depend on the benchmark scenario, the general strategy will be applicable more generally.

Regarding the first question, we found that in the presence of multiple axions, enhanced sensitivity is not always guaranteed. In LSW experiments, destructive interference in the photon and multiple axion oscillation system can suppress, or even eliminate, the regenerated photon signal for specific model and experimental parameter combinations.
However, such cancellations are typically tuned, and signals reappear  even for modest changes to the experimental parameters, such as the length of the magnetic regions or the wall thickness of an LSW experiment. For helioscopes, similar reductions may arise in special cases involving nearly degenerate axion spectra. For haloscopes, sensitivity depends critically on how the local DM density is distributed among the different axion species. While for fixed DM density, individual axion mode contributions are generically diluted, the presence of multiple resonant frequencies can increase the possibility of encountering a detectable signal. A detailed assessment of such interplays calls for combined analyses of cosmological production scenarios with experimental search strategies.

Going beyond an initial potential detection, a key objective would be to determine whether an observed signal originates from a single axion or from multiple axion species. In LSW experiments, this can be pursued by varying experimental parameters such as the production and regeneration region lengths, the wall thickness or the effective optical depth through a buffer gas. Measurements across different configurations can, in principle, be combined to reconstruct features of the underlying axion mass spectrum. In helioscopes, complementary information might be obtained by exploiting spectral or in-medium effects, while in haloscopes, the appearance of multiple resonant peaks would provide a clear indication of axion multiplicity. In all cases, careful search strategies should be developed accounting for practical limitations, such as resolution and scan coverage of the experiments. Nevertheless, even limited variations of experimental conditions can already provide valuable qualitative information and potential hints regarding a non-trivial origin of a signal and relation to multiple axions.

Our results broadly demonstrate that axion searches can probe not only particle masses and couplings, but also the underlying theory structure of the axion sector. Controlling experimental parameters affecting interference effects and coherence properties can render laboratory experiments sensitive to multi-field axion dynamics, effectively transforming them into spectroscopic probes of axion multiplicity. Taken together, our findings demonstrate that multiple axion scenarios can qualitatively modify experimental signatures and that existing and near-future experiments can possess diagnostic power to probe axion multiplicity, opening new avenues to explore the structure of axion sectors beyond the minimal single field paradigm.

\begin{acknowledgments}
This work was supported by World Premier International Research Center Initiative (WPI), MEXT, Japan. V.T. acknowledges support by the JSPS KAKENHI grant No.~23K13109.
This article/publication is based upon work from COST Action COSMIC WISPers CA21106, supported by COST (European Cooperation in Science and Technology).
A.d.G. thanks V.~Takhistov and the International Center for Quantum-field Measurement Systems for Studies of the Universe and Particles (QUP/KEK) for their hospitality and the stimulating working environment during which a core part of this work was realised. JJ would like to thank A.~Hebecker for discussions.
\end{acknowledgments}

\appendix
\section{Eikonal derivation of axion-photon conversion}
\label{app:eikonal}

In this appendix, we provide details of derivation of the multiple axion oscillation results using the eikonal (WKB) method of ref.~\cite{Adler:2008gk}.

We consider the EoM of $N$ weakly coupled axion fields $a$ and the photon field component ($A_{||}$) aligned with a background magnetic field $\vec{B}$, following from the Lagrangian of eq.~\eqref{eq:Lagr}. To be general, we include the possibility of having an effective photon mass due to medium effects,
\begin{align}\label{eik_EOM}
    &\Box \begin{pmatrix}
        A_{||}\\
        a_n
    \end{pmatrix}+\begin{pmatrix}
        m^2_\gamma &-c_ng_{a\gamma}B\partial_t\\
        c_ng_{a\gamma}B\partial_t & m_a^2
    \end{pmatrix}\begin{pmatrix}
        A_{||}\\
        a_n
    \end{pmatrix}=0\,.
\end{align}
Here, $a_{n}$ denotes the vector of axions.

We work in an experimental LSW configuration where propagation occurs along the $x$-direction, with the external magnetic $B$-field transverse to it. We consider monochromatic waves of energy $\omega$ and adopt the standard eikonal (WKB) ansatz for solutions
\begin{align}\label{eik_ansatz}
    A_{||}=A^0_{||}e^{i(\chi(x)-\omega t )}, \qquad
    a_n=a^0_{n}e^{i(\theta_n(x)-\omega t)}\,.
\end{align}

Inserting this into eq.~\eqref{eik_EOM} yields,
\begin{align}
    &(\omega^2-m^2_\gamma-(\chi')^2+i\chi'')A^0_{||}+i\omega\beta_n a^0_n=0\\
    &(\omega^2-m_n^2-(\theta_n')^2+i\theta_n'')a^0_{n}-i\omega\beta_n A^0_{||}=0\,,  \nonumber
\end{align}
with $\beta_n=g^n_{a\gamma}B$.
The index $n$ is summed over in the first equation.  Solving perturbatively in $g_{a\gamma}B$ and neglecting the second derivative terms, which would be required to describe reflected wave amplitudes more accurately, analogous to the procedure in ref.~\cite{Adler:2008gk}, the first order equations become
\begin{align}
    \left(2\sqrt{\omega^2-m^2_\gamma}\chi'^{(1)}\exp{\left(i\sqrt{\omega^2-m^2_\gamma} x\right)}\right)A^0_{||}-i\omega\beta_n \exp{\left(i\sqrt{\omega^2-m_n^2} x\right)}a^0_n&=0\\
    \left(i\omega\beta_n\exp{\left(i\sqrt{\omega^2-m^2_\gamma} x\right)}\right)A^0_{||}+2\sqrt{\omega^2-m_n^2}\theta_n'^{(1)}\exp{\left(i\sqrt{\omega^2-m_n^2} x)\right)}a^0_n&=0  \,.\nonumber
\end{align}
Here, $\chi^{(1)}$ and $\theta_n^{(1)}$ denote the first order coefficients in the Taylor expansion of $\theta$ and $\chi$ with respect to $\beta_n$.

We can solve these equations for $\chi$ and $\theta_n$ and reinsert the results into the ansatz solution of eq.~\eqref{eik_ansatz}. Assuming the magnetic field to be constant in a given region, the result is~\cite{Adler:2008gk}
\begin{align}\label{eik_wavesol}
    A_{||}(t,x)& = \exp{\left(i\sqrt{\omega^2-m^2_\gamma}x- i\omega t\right)}\Biggl[A_{||}^0-\beta_n  a^0_{n}\frac{\omega\left(\sqrt{\omega^2-m^2_\gamma}+\sqrt{\omega^2-m_n^2}\right)}{(m_n^2-m_\gamma^2)\sqrt{\omega^2-m_\gamma^2}} \\
& \quad  \times\sin{\left(\left(\sqrt{\omega^2-m_n^2}-\sqrt{\omega^2-m_\gamma^2}\right)x/2\right)}  \exp{(i\left(\sqrt{\omega^2-m_n^2}-\sqrt{\omega^2-m^2_\gamma}\right)x/2})\Biggr]\nonumber\\
    \label{eq:a-equat}a_n(t,x)& = \exp{\left(i\sqrt{\omega^2-m_n^2}x-i\omega t\right)}\Biggl[\Biggl(a^0_n+\beta_nA^0_{||}\frac{\omega\left(\sqrt{\omega^2-m^2_\gamma}+\sqrt{\omega^2-m_n^2}\right)}{(m_n^2-m_\gamma^2)\sqrt{\omega^2-m_n^2}} \\
& \quad  \times\sin{\left(\left(\sqrt{\omega^2-m_n^2}-\sqrt{\omega^2-m_\gamma^2}\right)x/2\right)} \exp{(i\left(\sqrt{\omega^2-m^2_\gamma}-\sqrt{\omega^2-m_n^2}\right)x/2})\Biggr)\Biggr]\nonumber  .
\end{align}
As before, we sum over $n$ in the first equation. Up to this point the only assumption that was made in the derivation of eq.~\eqref{eik_wavesol} is the validity of the WKB approximation.

To compute physical detection probabilities in the different considered types of experiments  we need to evaluate the absolute value squared of the relevant expressions.

\subsection{Light shining through wall experiments}

We first consider LSW experiments in the near-vacuum limit, with $m_\gamma \rightarrow 0$ and assuming $\omega>m_n$.
Then, the exponents in eq.~\eqref{eik_wavesol} are entirely imaginary.
For generality, we retain a small real part of the photon dispersion relation and take the limit $m_{\gamma R}\to 0$ at the end.

The initial conditions for regeneration region at $L_1+L_w$ are determined solely by the axion components, as all photon fields are absorbed by the wall. Interpreting the $x$ in the first equation of eq.~\eqref{eik_wavesol} as $x-L_1-L_w$ and neglecting global phases, these initial conditions give for the photon component at the end of the regeneration region
\begin{align}\label{eik_photonamp}
    A_{||}(t,L_1+L_w+L_2)=&-\beta_na_{n}(L_1+L_w)\frac{\omega\left(\sqrt{\omega^2-m^2_\gamma}+\sqrt{\omega^2-m_n^2}\right)}{(m_n^2-m_\gamma^2)\sqrt{\omega^2-m_\gamma^2}}
    \\& \times\sin{\left(\left(\sqrt{\omega^2-m_n^2}-\sqrt{\omega^2-m_\gamma^2}\right)L_2/2\right)}\nonumber \\& \times\exp{\left(i\left(\sqrt{\omega^2-m_n^2}-\sqrt{\omega^2-m_\gamma^2}\right)L_2/2\right)} \,.\nonumber
\end{align}

In the region between the two magnetic domains there are no interactions among fields involved. Each axion mode therefore propagates freely and acquires a phase  $\exp{(iL_w(\sqrt{\omega^2-m_n^2}))}$.

We can determine the axion field configuration at $L_1$ as follows. Considering that there are no incident axions at $x = 0$, eq.~\eqref{eq:a-equat} yields
\begin{align}
    a_{n}(t,L_1)& = \beta_nA^0_{||}\frac{\omega\left(\sqrt{\omega^2-m^2_\gamma}+\sqrt{\omega^2-m_n^2}\right)}{(m_n^2-m_\gamma^2)\sqrt{\omega^2-m_\text{n}^2}}\sin{\left(\left(\sqrt{\omega^2-m_n^2}-\sqrt{\omega^2-m_\gamma^2}\right)L_1/2\right)} \\
& \quad  \times\exp{\left(i\left(\sqrt{\omega^2-m_n^2}-\sqrt{\omega^2-m_\gamma^2}\right)L_1/2\right)}\,.\nonumber
\end{align}

Multiplying the respective axion components by the phase discussed earlier and inserting   into eq.~\eqref{eik_photonamp}  we obtain the final photon amplitude. Ultimately, we take the absolute value squared and normalize $A^0_{||}=1$ to obtain the detection probability,
\begin{align}\label{eik_prob}
P_\text{det.}=&\sum_n \frac{\beta_n^4 \omega^4\left(\sqrt{\omega^2-m_\gamma^2}+\sqrt{\omega^2-m_n^2}\right)^4}{(m_n^2-m^2_\gamma)^4(\omega^2-m_n^2)(\omega^2-m_\gamma^2)}\sin{\left(\left(\sqrt{\omega^2-m_n^2}-\sqrt{\omega^2-m_\gamma^2}\right)L_1/2\right)}
\\
& \times \sin{\left(\left(\sqrt{\omega^2-m_n^2}-\sqrt{\omega^2-m_\gamma^2}\right)L_2/2\right)} \nonumber
\\
&+\sum_{n<m}2\frac{\beta_n^2\beta_m^2 \omega^4\left(\sqrt{\omega^2-m_\gamma^2}+\sqrt{\omega^2-m_n^2}\right)^2\left(\sqrt{\omega^2-m_\gamma^2}+\sqrt{\omega^2-m_m^2}\right)^2}{(m^2_n-m^2_\gamma)^2(m^2_m-m^2_\gamma)^2\sqrt{\omega^2-m_n^2}\sqrt{\omega^2-m_m^2}(\omega^2-m_\gamma^2)}
\nonumber
\\
& \times\sin{\left(\left(\sqrt{\omega^2-m_n^2}-\sqrt{\omega^2-m_\gamma^2}\right)L_1/2\right)}\sin{\left(\left(\sqrt{\omega^2-m_m^2}-\sqrt{\omega^2-m_\gamma^2}\right)L_1/2\right)}\nonumber
\\ & \times \sin{\left(\left(\sqrt{\omega^2-m_n^2}-\sqrt{\omega^2-m_\gamma^2}\right)L_2/2\right)}\sin{\left(\left(\sqrt{\omega^2-m_m^2}-\sqrt{\omega^2-m_\gamma^2}\right)L_2/2\right)}
\nonumber
\\
& \times\cos{\left(\left(\sqrt{\omega^2-m_n^2}-\sqrt{\omega^2-m_m^2}\right)(L_1+2L_w+L_2)/2\right)}\,.\nonumber
\end{align}

In the limit $\omega\gg m_n$ and for the photon dispersion relation $\omega=k$ this reduces to
\begin{align}
        P_{\gamma\to X\to \gamma} & \simeq\left|\sum\limits_{n=1} \frac{4\kappa_n^2}{m_n^4}\sin\left(\frac{m_n^2L_1}{4\omega}\right)\sin\left(\frac{m_n^2L_2}{4\omega}\right)e^{-im_n^2(L_1+2L_w+L_2)/(4\omega)]}\right|^2
        \\
        &\nonumber = \left[ \sum\limits_{n=1} P_{\gamma\to n}(L_1)P_{\gamma\to n}(L_2)+2\sum\limits_{m<n} \Pi_{1,k}^\star\Pi_{2,k}^\star\Pi_{1,n}\Pi_{2,n}\cos\left(\frac{m_n^2-m_m^2}{4\omega}(L_1+2L_w+L_2)\right) \right].
\end{align}

\subsection{Helioscopes}

We now consider helioscopes. The quantity of interest is the probability for an axion to convert into a photon in the magnetic regeneration region of the helioscope. This can be obtained directly from eq.~\eqref{eik_wavesol} by setting the incident photon amplitude $A_{||}^0=0$ and taking the absolute value squared of the resulting photon field.

A fully rigorous prediction for the observable photon flux requires integrating this conversion probability over both the axion energy spectrum and the spatial production region inside the Sun. This procedure leads to eq.~\eqref{hel::prob} in the main text. For completeness,  we reproduce the corresponding expression here without taking the relativistic limit
\begin{align}\label{hel::probfull}
P=\;&\sum_n \eta_n^2 \frac{\beta_n^2 \omega ^2(\omega+\sqrt{\omega^2-m_n^2)})^2}{m_n^4(\omega^2-m_n^2)}\sin^{2}\biggl(\left(L_\text{mag}\left(\sqrt{\omega^2-m_n^2}-\omega\right)/2\right)\\&+\sum_{n\neq m}\eta_n\eta_m\frac{\beta_n\beta_m\omega^2(\omega+\sqrt{\omega^2-m_n^2})(\omega+\sqrt{\omega^2-m_m^2})}{m_n^2m_m^2\sqrt{\omega^2-m_n^2}\sqrt{\omega^2-m_m^2}}\sin\left(L_\text{mag}\left(\sqrt{\omega^2-m_n^2}-\omega\right)/2\right)\nonumber\\&\times\sin\left(L_\text{mag}\left(\sqrt{\omega^2-m_m^2}-\omega\right)/2\right)\cos\left((L_\text{Sun}+L_\text{mag}/2)\left(\sqrt{\omega^2-m_n^2}-\sqrt{\omega^2-m_m^2}\right)\right)\biggr)\,. \nonumber
\end{align}

In helioscope experiments, the characteristic axion energies lie in the keV range. In this regime, photons propagating through a buffer gas such as He acquire an effective mass. As discussed in the main text, varying the gas pressure and possibly temperature in the regeneration region, therefore, allows one to scan over effective photon masses and potentially achieve resonance.

 Technically, the exponent in eq.~\eqref{eik_wavesol} needs to be separated into its real and imaginary parts when computing the absolute value squared and evaluating the result at $x=L$. The resulting expressions simplify considerably in the large-$\omega$ limit, appropriate for helioscope searches and the bulk of the solar axion spectrum, and were given in the main text in eq.~\eqref{eq:P_medium}. The full expressions are provided here for completeness. The real $\kappa=\text{Re}(\sqrt{\omega^2-m_\gamma^2})$ and imaginary part $\delta=\text{Im}(\sqrt{\omega^2-m_\gamma^2})$ are given as
\begin{align}
    &\kappa=\left((\omega^2-m^2)^2+\omega^2\Gamma^2\right)^{1/4} \left(\frac{1+\frac{k^2_\gamma}{\Lambda}}{2}\right)^{1/2},  &&
    \delta= \left((\omega^2-m^2)^2+\omega^2\Gamma^2\right)^{1/4}\left(\frac{1-\frac{k^2_\gamma}{\Lambda}}{2}\right)^{1/2}\,, \nonumber \nonumber \\
\end{align}
with
    \begin{align}
    &k_\gamma^2= \omega^2-m^2_{\gamma R} , &&
    \Lambda^2=(\omega^2-m_{\gamma R}^2)^2+(\omega\Gamma)^2  .
\end{align}
Thus, for the axion-to-photon conversion probability, one obtains,
\begin{align}\label{eq:P_medium_app}
    &P_{a\rightarrow\gamma}=|K|^2 \left[\sin\left(\frac{\sqrt{\omega^2-m_n^2}-\kappa}{2} L\right)^2\exp(-\delta L)+\left(\frac{1-\exp(-\delta L)}{2}\right)^2\right]
    \end{align}
    with
    \begin{equation}
     |K|^2=(\omega\beta)^2   \left({(\kappa^2+\delta^2)\left(\left(\sqrt{\omega^2-m_n^2}-\kappa\right)^2+\delta^2\right)}\right)^{-1}\,
\end{equation}
Resonant enhancement occurs when $\sqrt{\omega^2-m_n^2}\simeq\kappa$.

\subsection{Conversion probability for superpositions with random phases}

We consider that at a fixed position inside the Sun, a superposition of axions is produced from two independent photon conversion events. The resulting axion field configurations have equal amplitudes and identical orientations in field space, but carry independent and random overall phases $\exp(i\eta^{a/b})$.

Accounting for these initial phases, the photon amplitude at the detector location can be written as
\begin{align}
    A(L_\text{Sun}+L_\text{mag})\simeq &-\sum_{n,a/b}\beta_n  \exp{(i L_\text{Sun}k_n+i\eta^{a/b})}\frac{\omega(k_n+\omega)}{m_n^2k_n} \notag\\&\times\sin{\left(((k_n-\omega) L_\text{mag})/2\right)} \exp{\left((i(k_n-\omega) L_\text{mag})/2\right)}\,.
\end{align}
Taking the absolute value squared, we find
\begin{align}
|A|^2\simeq\;&2\sum_n\frac{\beta_n^2\omega^2(k_n+\omega)^2}{k_n^2 m_n^4} \sin{\left(((k_n-\omega) L_\text{mag})/2\right)}^2
\\
& +\sum_{a/b} 2\frac{\beta_n^2\omega^2(k_n+\omega)^2}{k_n^2 m_n^4} \sin{\left(((k_n-\omega) L_\text{mag})/2\right)}^2 \cos{(\eta_a-\eta_b)} \nonumber
\\
&+ 4\sum_{n<m} \frac{\beta_n\beta_m\omega^2(k_n+\omega)(k_m+\omega)}{k_n k_m m_n^2 m_m^2} \sin{\left(((k_n-\omega) L_\text{mag})/2\right)}\sin{\left(((k_m-\omega) L_\text{mag})/2\right)} \nonumber
\\
& \times\cos{((L_\text{Sun}+L_\text{mag}/2)(k_n-k_m)})\nonumber
\\
& +4\sum_{n<m, a/b} \frac{\beta_n\beta_m\omega^2(k_n+\omega)(k_m+\omega)}{k_n k_m m_n^2 m_m^2} \sin{\left(((k_n-\omega) L_\text{mag})/2\right)}\sin{\left(((k_m-\omega) L_\text{mag})/2\right)} \nonumber
\\&
 \times\cos{(\eta_a-\eta_b)}\cos{((L_\text{Sun}+L_\text{mag}/2)(k_n-k_m)})\,. \nonumber
\end{align}
Averaging over the random relative phase $(\eta_a-\eta_b)$, the terms proportional to $\cos(\eta_a-\eta_b)$ vanish. The remaining contributions reproduce the structure of eq.~\eqref{hel::prob}, demonstrating that incoherent production leads to the same effective conversion probability as considered in the main text.

\bibliographystyle{BiblioStyle}
\bibliography{Bibliography}
\end{document}